\documentclass[final,3p,times,twocolumn]{elsarticle}

\usepackage{lineno}
\modulolinenumbers[5]

\usepackage[T1]{fontenc}
\usepackage{lmodern}

\usepackage[colorlinks,linkcolor=blue,citecolor=black,pdfpagelabels=true,pagebackref=true,plainpages=false]{hyperref}

\usepackage{todonotes}
\usepackage{booktabs}
\usepackage{siunitx}

\usepackage[toc,footnote,acronym]{glossaries}
\makeglossaries
\glsdisablehyper

\makeatletter
\newcommand*{\textoverline}[1]{$\overline{\hbox{#1}}\m@th$}
\makeatother

\usepackage[T1]{fontenc}
\usepackage[utf8]{inputenc}
\usepackage{lmodern}
\usepackage{textcomp}
\renewcommand{\ae}{ä}

\renewcommand{\oe}{ö}

\newcommand{\ue}{ü}

\newcommand{\executeiffilenewer}[3]{%
	\ifnum\pdfstrcmp{\pdffilemoddate{#1}}%
	{\pdffilemoddate{#2}}>0%
	{\immediate\write18{#3}}\fi%
}

\usepackage{siunitx}
\usepackage{eurosym}
\DeclareSIUnit\EUR{\text{\euro}}

\DeclareRobustCommand{\rxb}{\ensuremath{\vec{R}\times\vec{B}}}

\DeclareRobustCommand{\nomos}{NoMoS}

\def\TReg{\textsuperscript{\textregistered}}

\DeclareRobustCommand{\side}[1]{#1\nobreakdash-side}
\DeclareRobustCommand{\type}[1]{#1\nobreakdash-type}
\DeclareRobustCommand{\pnjunction}{p\nobreakdash-n\nobreakdash-junction}
\DeclareRobustCommand{\pstop}{p\nobreakdash-stop}

\DeclareRobustCommand{\apv}{APV25}

\DeclareRobustCommand{\CO2}{\ensuremath{\mathit{\mathrm{CO}}_2}}
\DeclareRobustCommand{\Btwo}{Belle~II}
\DeclareRobustCommand{\origami}{Origami}
\DeclareRobustCommand{\anti}{anti-}

\DeclareRobustCommand{\kB}{\ensuremath{k_\mathrm{B}}}

\DeclareRobustCommand{\xsuby}[2]{\ensuremath{#1_\mathrm{#2}}}

\DeclareRobustCommand{\beta}{\ensuremath{\upbeta}}

\newglossaryentry{backscattering}{
	name=backscattering,
	description={Backscattering (or backscatter) is the reflection of waves, particles, or signals back to the direction from which they came.
	It is usually a diffuse reflection due to scattering, as opposed to specular reflection as from a mirror, although specular backscattering can occur at normal incidence with a surface.
	Backscattering has important applications in astronomy, photography, medical ultrasonography and radiation sensor physics}}

\newglossaryentry{shot noise}{
	name=shot noise,
	description={Shot noise or Poisson noise is a type of noise which can be modeled by a Poisson process.
	In electronics shot noise originates from the discrete nature of electric charge.
	Shot noise also occurs in photon counting in optical devices, where shot noise is associated with the particle nature of light}}

\newglossaryentry{detection efficiency}{
	name=detection efficiency,
	description={The detection efficiency of a sensor is defined as the ratio between the number of detected particles and the number of incoming particles hitting the \gls{active area}, usually expressed in percent.
	Not to be confused with the ``counting efficiency'' of radiation detectors (ratio between the number of particles or photons counted with a radiation counter and the number of particles or photons of the same type and energy emitted by the radiation source) or with the ``quantum efficiency'' of photodetectors (ratio between the number of charge carriers collected at either terminal and the number of photons hitting the device's photoreactive surface)}}

\newglossaryentry{planar}{
	name=planar,
	description={Planar processing --- in contrast to 3D or monolithic processing --- only structures the surface of a semiconductor device in 2D, without creating structures in the bulk.
	The traditional planar silicon sensors are made this way, e.g., single-sided \glspl{sms}, \glspl{DSSD} and hybrid pixel sensors}}

\newglossaryentry{field stop}{
	name=field stop,
	description={The field stop is a heavily \glslink{doping}{doped} region at the sensor surface.
	Its high \gls{doping} concentration keeps the \gls{depletion} zone away from the surface}}

\newglossaryentry{recombination}{
	name=recombination,
	description={Recombination is the process by which an \gls{ehpair} is eliminated.
	It is facilitated by crystal defects, which introduce additional states in the \gls{band gap}}}

\newglossaryentry{multiplication layer}{
	name=multiplication layer,
	description={In semiconductor sensors, the multiplication layer is a deep, moderately strong \gls{implantation} at the \gls{pnjunction} with the same polarity as the \gls{substrate}. It increases the change rate of the effective \gls{doping} concentration at the \gls{pnjunction}, thus creating a peak in the electric field. \Glspl{electron} traversing this field peak experience \gls{charge multiplication}}}

\newglossaryentry{impact ionization}{
	name=impact ionization,
	description={Impact ionization is the process in a material by which one energetic charge carrier can lose energy by the creation of other charge carriers. For example, in semiconductors, an \gls{electron} (or \gls{hole}) with enough kinetic energy can knock a bound \gls{electron} out of its bound state (in the \gls{valence band}) and promote it to a state in the \gls{conduction band}, creating an \gls{ehpair}. For carriers to have sufficient kinetic energy a sufficiently large electric field must be applied, in essence requiring a sufficiently large voltage but not necessarily a large current. If this occurs in a region of high electrical field then it can result in avalanche breakdown. This process is exploited in \glspl{APD}, by which a small optical signal is amplified before entering an external electronic circuit}}

\newglossaryentry{charge multiplication}{
	name=charge multiplication,
	description={Charge multiplication is a process in which a number of free \glspl{electron} in a transmission medium are subjected to strong acceleration by an electric field and subsequently collide with other atoms of the medium, thereby ionizing them (\gls{impact ionization}). This releases additional \glspl{electron} which accelerate and collide with further atoms, releasing more \glspl{electron} --- a chain reaction. Also known as ``electron avalanche'' or ``Townsend discharge''}}

\newglossaryentry{internal gate}{
	name=internal gate,
	description={The internal gate is a feature of a \gls{DEPFET} pixel. It is a potential minimum in the \gls{bulk} of the sensor, created by deep \gls{implantation}. It is located underneath the ``external gate'' of the \gls{FET}, which is integrated into the pixel. The potential minimum works as a trap for signal \glspl{electron}. The amount of charge stored in the internal gate modulates the current of the \gls{FET}. See~\cite{Kemmer1987} for details}}

\newglossaryentry{polarity}{
	name=polarity,
	plural=polarities,
	description={Electrical polarity is a term used throughout industries and fields that involve electricity. There are two types of poles: positive (+) and negative (-). This represents the electrical potential at the ends of a circuit. A battery has a positive terminal (+ pole) and a negative terminal (- pole). Interconnection of electrical device nearly always require correct polarity to be maintained. Correct polarity is essential for the operation of vacuum tube and semiconductor devices, many electric motors, electrochemical cells, electrical instruments, and other devices}}

\newglossaryentry{substrate}{
	name=substrate,
	description={Substrate is a term used in materials science to describe the base material on which processing is conducted to produce new film or layers of material such as deposited coatings}}

\newglossaryentry{veto}{
	name=veto,
	description={A veto (Latin for ``I forbid'') is the power to stop an action.
	In particle physics, veto detectors are used to mark measurement events as faulty, if the veto detector saw a signal.
	This is done to suppress the influence of background events or non-regular states in the system}}

\newglossaryentry{COMSOL}{
	name=COMSOL,
	description={COMSOL Multiphysics is a cross-platform finite element analysis, solver and multiphysics simulation software.
	It allows conventional physics-based user interfaces and coupled systems of partial differential equations.
	COMSOL provides an \gls{IDE} and unified workflow for electrical, mechanical, fluid, acoustics and chemical applications}}

\newglossaryentry{beta decay}{
	name=beta decay,
	description={In nuclear physics, beta decay ($\beta$-decay) is a type of radioactive decay in which a beta particle (fast energetic \gls{electron} or \gls{positron})
	is emitted from an atomic nucleus.
	For example, beta decay of a \gls{neutron} transforms it into a \gls{proton} by the emission of an \gls{electron} accompanied by an anti\gls{neutrino},
	or conversely a \gls{proton} is converted into a \gls{neutron} by the emission of a \gls{positron} (\gls{positron} emission) with a \gls{neutrino}, thus changing the nuclide type.
	Neither the beta particle nor its associated (anti-)\gls{neutrino} exist within the nucleus prior to beta decay, but are created in the decay process}}

\newglossaryentry{Hamiltonian}{
	name=Hamiltonian,
	description={In quantum mechanics, a Hamiltonian is an operator corresponding to the sum of the kinetic energies plus the 
	potential energies for all the particles in the system (this addition is the total energy of the system in most of the cases under analysis).
	It is usually denoted by $H$ $\check{H}$ or $\hat{H}$.
	Its spectrum is the set of possible outcomes when one measures the total energy of a system.
	Because of its close relation to the time-evolution of a system, it is of fundamental importance in most formulations of quantum theory.
  The Hamiltonian is named after William Rowan Hamilton, who created a revolutionary reformulation of Newtonian mechanics,
	now called Hamiltonian mechanics, which is also important in quantum physics}}

\newglossaryentry{perturbation theory}{
	name=perturbation theory,
	description={In quantum mechanics, perturbation theory is a set of approximation schemes directly related to mathematical perturbation for 
	describing a complicated quantum system in terms of a simpler one.
	The idea is to start with a simple system for which a mathematical solution is known, and add an additional ``perturbing'' \gls{Hamiltonian} representing a 
	weak disturbance to the system.
	If the disturbance is not too large, the various physical quantities associated with the perturbed system (e.g., its energy levels and eigenstates)
	can be expressed as ``corrections'' to those of the simple system.
	These corrections, being small compared to the size of the quantities themselves, can be calculated using approximate methods such as asymptotic series.
	The complicated system can therefore be studied based on knowledge of the simpler one.
	In effect, it is describing a complicated unsolved system using a simple, solved system}}

\newglossaryentry{Lie group}{
	name=Lie group,
	description={In mathematics, a Lie group is a group that is also a differentiable manifold, with the property that the group operations are smooth.
	Lie groups are named after Norwegian mathematician Sophus Lie, who laid the foundations of the theory of continuous transformation groups.
  In rough terms, a Lie group is a continuous group, that is, one whose elements are described by several real parameters.
	As such, Lie groups provide a natural model for the concept of continuous symmetry, such as rotational symmetry in three dimensions.
	Lie groups are widely used in many parts of modern mathematics and physics}}

\newglossaryentry{Lagrangian}{
	name=Lagrangian,
	description={Lagrangian field theory is a formalism in classical field theory.
	It is the field-theoretic analogue of Lagrangian mechanics.
	Lagrangian mechanics is used for discrete particles each with a finite number of degrees of freedom.
	Lagrangian field theory applies to continua and fields, which have an infinite number of degrees of freedom.
	In Lagrangian field theory, the Lagrangian as a function of generalized coordinates is replaced by a Lagrangian density,
	a function of the fields in the system and their derivatives, and possibly the space and time coordinates themselves.
	In field theory, the independent variable $t$ is replaced by an event in spacetime ($x$, $y$, $z$, $t$) or still more generally by a point $s$ on a manifold.
  Often, a ``Lagrangian density'' is simply referred to as a ``Lagrangian''}}

\newglossaryentry{gauge theory}{
	name=gauge theory,
	description={In physics, a gauge theory is a type of field theory in which the \gls{Lagrangian} is invariant under certain \glspl{Lie group} of local transformations.
The term ``gauge'' refers to any specific mathematical formalism to regulate redundant degrees of freedom in the \gls{Lagrangian}.
The transformations between possible gauges, called gauge transformations, form a \gls{Lie group} --- referred to as the symmetry group or the gauge group of the theory.
Associated with any \gls{Lie group} is the Lie algebra of group generators.
For each group generator there necessarily arises a corresponding field (usually a vector field) called the gauge field.
Gauge fields are included in the \gls{Lagrangian} to ensure its invariance under the local group transformations (called gauge invariance).
When such a theory is quantized, the quanta of the gauge fields are called gauge \glspl{boson}.
Many powerful theories in physics are described by \glspl{Lagrangian} that are invariant under some symmetry transformation groups.
When they are invariant under a transformation identically performed at every point in the spacetime in which the physical processes occur,
they are said to have a global symmetry.
Local symmetry, the cornerstone of gauge theories, is a stronger constraint.
In fact, a global symmetry is just a local symmetry whose group's parameters are fixed in spacetime
(the same way a constant value can be understood as a function of a certain parameter, the output of which is always the same).
Gauge theories are important as the successful field theories explaining the dynamics of elementary particles}}

\newglossaryentry{hierarchy problem}{
	name=hierarchy problem,
	description={In theoretical physics, the hierarchy problem is the large discrepancy between aspects of the \gls{weak} force and \gls{gravitation}.
	There is no scientific consensus on why, for example, the \gls{weak} force is \SI{1e24}{} times stronger than \gls{gravitation}}}

\newglossaryentry{Hall probe}{
	name=Hall probe,
	description={A Hall probe (or Hall effect sensor) is a device that is used to measure the magnitude of a magnetic field.
	Its output voltage is directly proportional to the magnetic field strength through it.
Hall effect sensors are used for proximity sensing, positioning, speed detection, and current sensing applications.
In a Hall effect sensor, a thin strip of metal has a current applied along it.
In the presence of a magnetic field, the \glspl{electron} in the metal strip are deflected toward one edge,
producing a voltage gradient across the short side of the strip (perpendicular to the feed current).
Hall effect sensors have an advantage over inductive sensors in that,
while inductive sensors respond to a changing magnetic field which induces current in a coil of wire and produces voltage at its output,
Hall effect sensors can detect static (non-changing) magnetic fields}}

\newglossaryentry{systematic}{
	name=systematic,
	description={A systematic error is predictable and typically constant or proportional to the true value, the so-called systematic parameter.
	If the cause of the systematic error can be identified, then it usually can be eliminated.
	Systematic errors are caused by imperfect calibration of measurement instruments or imperfect methods of observation,
	or interference of the environment with the measurement process, and always affect the results of an experiment in a predictable direction.
	Incorrect zeroing of an instrument leading to a zero error is an example of systematic error in instrumentation}}

\newglossaryentry{azimuth}{
	name=azimuth,
	description={An azimuth is an angular measurement in a spherical coordinate system.
	The vector from an observer (origin) to a point of interest is projected perpendicularly onto a reference plane;
	the angle between the projected vector and a reference vector on the reference plane is called the azimuth}}

\newglossaryentry{gyration}{
	name=gyration,
	description={Gyration is the circular motion of a charged particle in the presence of a uniform magnetic field}}

\newglossaryentry{transport function}{
	name=transport function,
	description={The transport function encompasses our whole knowledge of the \gls{nomos} system.
It allows us to calculate the \gls{electron} and \gls{proton} spectrum as seen by the \gls{drift detector} in form of a 2D histogram, including all \gls{systematic} effects.
See \cref{sec:transport_function} for more details}}

\newglossaryentry{post-acceleration}{
	name=post-acceleration,
	description={Post-acceleration is the process of accelerating particles after they have been created by the studied process.
	In our case, post-acceleration is necessary to make the \glspl{proton} from free \gls{neutron} decay detectable.
	Without post-acceleration, the \glspl{proton} have an energy below \SI{800}{\eV}, which is way below the detection threshold of contemporary detection techniques.
	We use a \gls{HV} electrode to post-accelerate the \glspl{proton} to \SIrange{10}{15}{\kilo\eV}}}

\newglossaryentry{Monte Carlo}{
	name=Monte Carlo,
	description={Monte Carlo methods, or Monte Carlo experiments, are a broad class of computational algorithms that rely on repeated random sampling to obtain numerical results.
	The underlying concept is to use randomness to solve problems that might be deterministic in principle.
	They are often used in physical and mathematical problems and are most useful when it is difficult or impossible to use other approaches.
	Monte Carlo methods are mainly used in three problem classes: optimization, numerical integration, and generating draws from a probability distribution}}

\newglossaryentry{rxb region}{
	name=\rxb{} region,
	description={The \rxb{} region is the part of the \gls{nomos} experiment, where we have a toroidal curved magnetic field.
	This is the region where the \gls{neutron}'s decay products experience a drift proportional to their momenta, perpendicular to both the magnetic field direction and to the curvature radius}}

\newglossaryentry{drift detector}{
	name=drift detector,
	description={The drift detector is mounted in the \gls{detection region} and measures the drift distances of the \gls{neutron}'s decay products.
	It will most likely be a position-resolved silicon sensor with a thin \gls{entrance window}}}
	
\newglossaryentry{entrance window}{
	name=entrance window,
	description={The entrance window of a detector is the part which admits the particles to be measured into the sensitive volume.
	In case of a silicon sensor the entrance window is one side of the sensor, which ideally is unstructured, homogeneous, and has very thin inactive layers.
	The inactive layers of a silicon sensor sometimes includes a metal layer, usually a silicon oxide layer,
	and always a highly \glslink{doping}{doped} layer inside the silicon \gls{substrate} itself}}
	
\newglossaryentry{detection region}{
	name=detection region,
	description={The detection region houses the \gls{drift detector}, which measures the drift distances of the decay products.
	To provide a homogeneous magnetic field, it includes a number of magnetic coils.
	For \gls{proton} measurements, we will put a \gls{HV} electrode in front of the \gls{drift detector} for \gls{post-acceleration}}}
	
\newglossaryentry{rail system}{
	name=rail system,
	description={We plan to install a rail system in the \gls{nomos} apparatus, which allows us to move a small car along the vacuum tube, especially in the \rxb{} region.
	This car will be equipped with magnetic field sensors.
	That way, we can gather a precise map of the magnetic field inside the system}}

\newglossaryentry{bWM}{
	name=weak magnetism form factor \xsuby{b}{WM},
	symbol=\xsuby{b}{WM},
	description={\xsuby{b}{WM} is also related to the \gls{electron} spectrum
	\todo{glossary.tex: write short description of the weak magnetism form factor}}}

\newglossaryentry{tensor coupling}{
	name=tensor coupling,
	description={\todo{glossary.tex: write short description of tensor coupling}}}

\newglossaryentry{B}{
	name=beta asymmetry parameter $B$,
	symbol=\ensuremath{B},
	description={The beta asymmetry parameter $B$ quantifies the correlation of \gls{neutrino} momentum and \gls{neutron} spin\todo{glossary.tex: extend description of B}}}

\newglossaryentry{A}{
	name=beta asymmetry parameter $A$,
	symbol=\ensuremath{A},
	description={The beta asymmetry parameter $A$ quantifies the correlation of \gls{electron} momentum and \gls{neutron} spin\todo{glossary.tex: extend description of A}}}

\newglossaryentry{lambda}{
	name=axial-vector to vector coupling constant $\lambda$,
	symbol=\ensuremath{\lambda},
	description={\todo{glossary.tex: write short description of the axial-vector to vector coupling constant $\lambda$}}}

\newglossaryentry{supersymmetry}{
name=supersymmetry,
description={In particle physics, supersymmetry (SUSY) is a principle that proposes a relationship between two basic classes of elementary particles:
\glspl{boson}, which have an integer-valued spin, and \glspl{fermion}, which have a half-integer spin.
A type of spacetime symmetry, supersymmetry is a possible candidate for undiscovered particle physics,
and seen as an elegant solution to many current problems in particle physics if confirmed correct,
which could resolve various areas where current theories are believed to be incomplete.
A supersymmetrical extension to the \gls{Standard Model} would resolve major \glspl{hierarchy problem} within \gls{gauge theory},
by guaranteeing that quadratic divergences of all orders will cancel out in \gls{perturbation theory}}}

\newglossaryentry{Higgs sector}{
	name=Higgs sector,
	description={In particle physics, the Higgs sector is the collection of quantum fields and/or particles that are responsible for the Higgs mechanism, i.e., for the spontaneous symmetry breaking of the Higgs field.
	The word ``sector'' refers to a subgroup of the total set of fields and particles}}

\newglossaryentry{aberration}{
	name=aberration,
	description={An aberration is a deviation from what is typical or normal}}

\newglossaryentry{adiabaticity}{
	name=adiabaticity,
	description={Adiabaticity is the condition of being adiabatic; a measure of the extent to which a process is adiabatic.
	Magnetic particle transport is adiabatic, when the particles ~\glslink{gyration}{gyrate} around a magnetic field line without leaving it. 
	Non-adiabatic effects can cause distortions of the position measurement and outright particle loss}}

\newglossaryentry{aperture}{
	name=aperture,
	description={The aperture is a special kind of collimator.
	Its primary purpose is to narrow the beam of \glspl{electron} and \glspl{proton} which enters the sensitive region of the instrument, i.e., to cause the spatial cross section of the beam to become smaller (beam limiting device).
	Furthermore, it will be equipped with sensors which deliver \gls{veto} signals from backscattered \glspl{electron}}}

\newglossaryentry{Doppler effect}{
	name=Doppler effect,
	description={The Doppler effect (or the Doppler shift) is the change in frequency or wavelength of a wave in relation to an observer who is moving relative to the wave source.
	It is named after the Austrian physicist Christian Doppler, who described the phenomenon in 1842.
	In our case, the Doppler effect changes the momenta of the \gls{neutron} decay products}}

\newglossaryentry{work function}{
	name=work function,
	description={In solid-state physics, the work function (sometimes spelled workfunction) is the minimum thermodynamic work (i.e., energy) needed to remove an \gls{electron} from a solid to a point in the vacuum immediately outside the solid surface.
	Here ``immediately'' means that the final \gls{electron} position is far from the surface on the atomic scale, but still too close to the solid to be influenced by ambient electric fields in the vacuum.
	The work function is not a characteristic of a \gls{bulk} material, but rather a property of the surface of the material (depending on crystal face and contamination)}}

\newglossaryentry{b}{
	name=Fierz interference term $b$,
	symbol=\ensuremath{b},
	description={The Fierz interference term $b$ describes a distortion of the \gls{electron} spectrum, which goes beyond the description of the \gls{Standard Model}.
	It can be determined from the \gls{electron} spectrum of free \gls{neutron} decay\todo{glossary.tex: extend description of b}}}
	
\newglossaryentry{a}{
	name=electron-neutrino correlation coefficient $a$,
	symbol=\ensuremath{a},
	description={The \gls{electron}-\gls{neutrino} correlation coefficient $a$ can be related to the \gls{proton} recoil spectrum.
	It can be determined from the \gls{proton} spectrum of free \gls{neutron} decay\todo{glossary.tex: extend description of a}}}
	
\newglossaryentry{alpha}{
	name=opening angle $\alpha$,
	symbol=\ensuremath{\alpha},
	description={The opening angle alpha is defined by the region of \gls{nomos}, in which the decay products experience a pure \rxb{} drift. 
	Nominally, $\alpha = \SI{180}{\degree}$ (for the nominal system) or $\alpha = \SI{90}{\degree}$ (for the reduced system).
	Unavoidable inhomogeneities will reduce $\alpha$ depending on the particle position}}

\newglossaryentry{helmholtz coil}{
	name=Helmholtz coil,
	description={A Helmholtz coil is a device for producing a region of nearly uniform magnetic field, named after the German physicist Hermann von Helmholtz.
	It consists of two electromagnets on the same axis}}

\newglossaryentry{decay volume}{
	name=decay volume,
	description={The decay volume is the region of the experiment, where free \glspl{neutron} can enter, and --- if they decay -- its decay products can enter the sensitive part of the apparatus. It can be subjected to a magnetic field by a pair of \glspl{helmholtz coil}}}

\newglossaryentry{magnetic filter}{
	name=magnetic filter,
	description={A magnetic filter is a magnetic field peak in the trajectory of a charged particle, which --- depending on the relation between the filter field and the surrounding field --- repels particles depending on their angle via the magnetic mirror effect.
	Particles with low angles can pass, while high angles are reflected}}

\newglossaryentry{hybrid}{
	name=hybrid,
	description={The hybrid board is an electronic board which houses readout chips, supplying power, control signals and routing their output to the subsequent data cables}}

\newglossaryentry{Belle}{
	name=Belle,
	description={Belle is the name of a former particle physics experiment at \gls{kek} (Tsukuba, Japan). Belle is not an acronym, so it's written with a capital letter at the beginning only. Belle is the french word for ``beautiful'', and hints to the b \gls{quark}, also called ``beauty \gls{quark}''}}

\newglossaryentry{BaBar}{
	name=BaBar,
	description={BaBar is the name of a particle physics experiment at the PEP-II collider at SLAC National Accelerator Laboratory, which was operated by Stanford University for the Department of Energy in California. It was a ``friendly competitor'' of the \gls{Belle} experiment, also measuring \gls{CP} violation with \glspl{B meson}}}

\newglossaryentry{BelleII}{
	name=Belle~II,
	description={\Btwo{} is the name of a future particle physics experiment at \gls{kek} (Tsukuba, Japan) which is presently under construction. It is the successor of the \gls{Belle} experiment}}

\newglossaryentry{quark}{
	name=quark,
	description={A quark is an elementary particle and a fundamental constituent of matter. Quarks combine to form composite particles called \glspl{hadron}, the most stable of which are \glspl{proton} and \glspl{neutron}, the components of atomic nuclei. The quarks of the first \gls{generation} are called ``up'' and ``down'', those of the second \gls{generation} are called ``charm'' and ``strange'', and those of the third \gls{generation} are called ``top'' and ``bottom'' (or ``beauty'')}}

\newglossaryentry{baryon}{
	name=baryon,
	description={A baryon is a composite subatomic particle made up of three \glspl{quark}}}

\newglossaryentry{meson}{
	name=meson,
	description={A meson is a composite subatomic particle composed of one \gls{quark} and one anti-quark}}

\newglossaryentry{hadron}{
	name=hadron,
	description={A hadron is a non-fundamental particle composed of \glspl{quark}. Hadrons with three \glspl{quark} are called \glspl{baryon}, hadrons with a quark and an anti-quark are called \glspl{meson}}}

\newglossaryentry{lepton}{
	name=lepton,
	description={A lepton is an fundamental particle which is assumed to be pointlike. The leptons of the \glslink{SM}{standard model} are the \gls{electron} $\mathrm{e}^-$, the \gls{muon} $\upmu^-$, the \gls{tau} $\uptau^-$, the corresponding \glspl{neutrino} $\upnu_\mathrm{e}$, $\upnu_\upmu$ and $\upnu_\uptau$, as well as their \anti particles. A lepton does not undergo \gls{strong} interactions, it interacts \glslink{em}{electromagnetically}, \glslink{weak}{weakly} and \glslink{gravitation}{gravitationally}. $\uplambda\upepsilon\uppi\uptau\acute{o}\upvarsigma$ (greek): small, thin, tiny}}

\newglossaryentry{photon}{
	name=photon,
	description={The photon $\upgamma$ is an elementary particle, the quantum of light and all other forms of electromagnetic radiation, and the mediator of the electromagnetic force, even when static via virtual photons}}

\newglossaryentry{proton}{
	name=proton,
	symbol=p,
	description={The proton (symbol: p or p$^+$) is a subatomic particle with one positive electric elementary charge. One or more protons are present in the nucleus of each atom. In the modern Standard Model of particle physics, the proton is a \gls{hadron}, composed of two up \glspl{quark} and one down \gls{quark}}}

\newglossaryentry{neutron}{
	name=neutron,
	symbol=n,
	description={The neutron (symbol: n or n$^0$) is a subatomic particle with no net electric charge and a mass slightly larger than that of a \gls{proton}. With the exception of \mbox{hydrogen-1}, nuclei of atoms consist of \glspl{proton} and neutrons, which are therefore collectively referred to as nucleons. In the modern Standard Model of particle physics, the neutron is a \gls{hadron}, composed of one up \gls{quark} and two down \glspl{quark}}}
	
\newglossaryentry{electron}{
	name=electron,
	symbol=e,
	description={The electron (symbol: e$^-$) is a subatomic particle with a negative elementary electric charge. It is a \gls{lepton} of the first \gls{generation} of fundamental particles}}

\newglossaryentry{muon}{
	name=muon,
	description={The muon (symbol: \textmu$^-$) is a subatomic particle with a negative elementary charge.  It is a \gls{lepton} of the second \gls{generation} of fundamental particles}}

\newglossaryentry{tau}{
	name=tau,
	description={The tau particle (symbol: $\uptau^-$) is a subatomic particle with a negative elementary charge.  It is a \gls{lepton} of the third \gls{generation} of fundamental particles}}
	
\newglossaryentry{strong}{
	name=strong,
	description={The strong interaction (or strong force) is the force that binds \glspl{proton} and \glspl{neutron} (nucleons) together to form the nucleus of an atom. It furthermore is the force (carried by gluons) that holds \glspl{quark} together to form \glspl{proton}, \glspl{neutron} and other \glspl{hadron}}}
	
\newglossaryentry{weak}{
	name=weak,
	description={The weak interaction is responsible for the radioactive decay of subatomic particles and initiates the process known as hydrogen fusion in stars. Weak interactions affect all known \glspl{fermion}. It is mediated by the exchange of the Z, W$^+$ and W$^-$ \glspl{boson}}}
	
\newglossaryentry{em}{
	name=electromagnetic,
	description={The electromagnetic interaction is the force that acts on electrically charged particles, allowing them to be accelerated and trapped in bound states. Thus, it is responsible for the interaction between the atomic nucleus and the \gls{electron} cloud. It is mediated by the \gls{photon}}}
	
\newglossaryentry{gravitation}{
	name=gravitation,
	description={Gravitation is the fourth fundamental interaction known today. It is the agent that gives weight to objects that have mass. Up to now it cannot be described with the same mathematical formalism as is used to describe the other three interactions. Therefore it could not yet be integrated into a consistent theory of all four interactions. Gravitation is described by the theory of general relativity, while the other interactions are described with quantum field theories}}
	
\newglossaryentry{neutrino}{
	name=neutrino,
	symbol=$\upnu$,
	description={The neutrino (symbol: $\upnu_\mathrm{e}$, $\upnu_\upmu$ or $\upnu_\uptau$, depending on the particle \gls{generation}) is an electrically neutral \gls{lepton}, which only interacts by the \gls{weak} interaction and by \gls{gravitation}. Is is therefore extremely difficult to detect}}

\newglossaryentry{generation}{
	name=generation,
	description={In particle physics, a generation (or family) is a division of the elementary particles. Between generations, particles differ by their (\gls{flavour}) quantum number and mass, but their interactions are identical.
There are three generations according to the Standard Model of particle physics. Each generation is divided into two \glspl{lepton} and two \glspl{quark}. The two leptons may be classified into one with electric charge -1 (\gls{electron}-like) and one neutral (\gls{neutrino}); the two quarks may be classified into one with charge -1/3 (down-type) and one with charge +2/3 (up-type). Every particle has a corresponding \anti particle}}
	
\newglossaryentry{fermion}{
	name=fermion,
	description={A fermion is a (not necessarily fundamental) particle with a half-integer spin quantum number. It obeys the Fermi-Dirac statistics and the Pauli exclusion principle, stating that no two fermions with an identical set of quantum numbers can occupy the same state in a system. All \glspl{lepton} and \glspl{quark} are fermions}}

\newglossaryentry{boson}{
	name=boson,
	description={A boson is a (not necessarily fundamental) particle with an integer spin quantum number. It obeys the Bose-Einstein statistics but not the Pauli exclusion principle. So, two bosons with an identical set of quantum numbers can occupy the same state in a system. All mediators of the fundamental interactions are bosons}}
	
\newglossaryentry{charge conjugation}{
	name=charge conjugation,
	description={The charge conjugation transforms a particle into its \anti particle by reversing all of its quantum numbers, especially the electrical charge}}
	
\newglossaryentry{charge-parity conjugation}{
	name=charge-parity conjugation,
	description={The charge-parity conjugation is the combination of the \gls{charge conjugation} and the \gls{parity conjugation}}}

\newglossaryentry{parity conjugation}{
	name=parity conjugation,
	description={The parity conjugation is a symmetry operation which mirrors a vector at its foot point}}
	
\newglossaryentry{Higgs}{
	name=Higgs,
	description={The Higgs \gls{boson} explains why some fundamental particles have mass when the symmetries controlling their interactions should require them to be massless, and -- related to this -- why the \gls{weak} force has a much shorter range than the \gls{em} force}}
	
\newglossaryentry{electro-weak}{
	name=electro-weak,
	description={The electro-weak interaction is the unified fundamental force that parents the \gls{em} interaction and the \gls{weak} interaction}}
	
\newglossaryentry{Standard Model}{
	name=Standard Model,
	description={The Standard Model of particle physics is a theory concerning the electromagnetic, weak, and strong nuclear interactions, which mediate the dynamics of the known subatomic particles. Because of its success in explaining a wide variety of experimental results, the Standard Model is sometimes regarded as a ``theory of almost everything''. However, the Standard Model falls short of being a complete theory of fundamental interactions because it makes certain simplifying assumptions. It does not incorporate the full theory of gravitation as described by general relativity, or predict the accelerating expansion of the universe (as possibly described by dark energy). It also does not correctly account for neutrino oscillations (and their non-zero masses)}}

\newglossaryentry{solenoid}{
	name=solenoid,
	description={A solenoid is a coil wound into a tightly packed helix. Such a coil produces a very homogenous magnetic field in its center}}

\newglossaryentry{eV}{
	name=eV,
	description={The ``electronvolt'' is the commonly used energy unit in particle physics. $1\,\mathrm{eV} \approx 1.6 \times 10^{-19}\,\mathrm{J}$ is the energy an \gls{electron} gains when it is accelerated by a potential difference of 1\,Volt. It is often used as 1\,GeV = $10^9$\,eV. Derived units are eV/c for the momentum, and eV/c$^2$ for the mass}}
	
\newglossaryentry{B factory}{
	name=B~factory,
	description={A B factory is a particle collider specifically designed to deliver a huge amount of \glspl{B meson}}}	
	
\newglossaryentry{B meson}{
	name=B~meson,
	description={B \glspl{meson} contain a \textoverline{b} \anti \gls{quark}, whereas B \anti \glspl{meson} contain a b~\gls{quark}. Those quarks are accompanied by u or d~quarks or their \anti particles, respectively}}

\newglossaryentry{positron}{
	name=positron,
	description={The positron (symbol: e$^+$) is a subatomic particle with a positive elementary electric charge. It is the \anti particle of the \gls{electron}}}

\newglossaryentry{quark mixing}{
	name=quark mixing,
	description={Quark mixing describes the mismatch of quantum states of \glspl{quark} when they propagate freely and when they take part in the \gls{weak} interactions. The \gls{quark} mass eigenstates are not the same as the eigenstates of the \gls{weak} interaction, but linear combinations of them. This is considered by the \gls{CKM matrix}}}

\newglossaryentry{unitary}{
	name=unitary,
	description={A complex square matrix $U$ is unitary if $U^*U = UU^* = I$, where $I$ is the identity matrix and $U^*$ is the conjugate transpose of $U$. The real analogon of a unitary matrix is an orthogonal matrix}}

\newglossaryentry{eigenstate}{
	name=eigenstate,
	description={The word ``eigenstate'' is derived from the German/Dutch word ``eigen'', meaning ``inherent'' or ``characteristic''. An eigenstate is the measured state of some object possessing quantifiable characteristics such as position, momentum, etc. The state being measured and described must be observable (i.e.\ something such as position or momentum that can be experimentally measured either directly or indirectly), and must have a definite value, called an eigenvalue}}	

\newglossaryentry{flavour}{
	name=flavour,
	description={Flavour refers to the type of elementary particles (either quarks or leptons) occurring in the \glslink{SM}{Standard Model}. There are flavour quantum numbers which depend on the number of particles of particular flavours which occur in a hadron. In \gls{strong} interactions, flavour is conserved. In the \gls{weak} interaction, however, this symmetry is broken, and flavour changing processes exist, such as \gls{quark} decay or \gls{neutrino} oscillations}}	

\newglossaryentry{B physics}{
	name=B~physics,
	description={B physics deals with the interactions and phenomena found in \gls{B meson} systems, especially \gls{CP} violation}}	

\newglossaryentry{kaon}{
	name=kaon,
	description={A kaon (symbol: K) is any of a group of four \glspl{meson} distinguished by a quantum number called strangeness. In the quark model they are understood to be bound states of a strange \gls{quark} (s) (or \anti quark \textoverline{s}) and an up or down \anti quark (\textoverline{u} or \textoverline{u}) (or quark u or d). Although the neutral kaon K$^0$ and its \anti particle \textoverline{K}$^0$ are usually produced via the \gls{strong} force, they decay \glslink{weak}{weakly}. Thus, once created the two are better thought of as superpositions of two \gls{weak} \glspl{eigenstate} which have vastly different lifetimes: The long-lived neutral kaon is called the K$_\mathrm{L}$ (``K-long'') and the short-lived neutral kaon is called the K$_\mathrm{S}$ (``K-short'')}}

\newglossaryentry{tree level}{
	name=tree level,
	description={In theoretical particle physics, tree level denotes the $0^\mathrm{th}$ iteration of a \gls{perturbation calculation}}}	
	
\newglossaryentry{perturbation calculation}{
	name=perturbation calculation,
	description={In quantum mechanics, perturbation theory is a set of approximation schemes directly related to mathematical perturbation for describing a complicated quantum system in terms of a simpler one}}	

\newglossaryentry{upsilon}{
	name=$\Upsilon$,
	description={The $\Upsilon$ \gls{meson} is a so-called ``bottonium'', consisting of a b~\gls{quark} and of a \textoverline{b} \anti \gls{quark}}}	

\newglossaryentry{vertex}{
	name=vertex,
	description={A vertex is the common origin of two or more particle trajectories},
	plural=vertices}	

\newglossaryentry{b-tagging}{
	name=b-tagging,
	description={b-tagging is the process of identifying a \gls{B meson} from its decay products, and determining if it contained a b \gls{quark} or a \textoverline{b} \anti \gls{quark}}}
	
\newglossaryentry{branching fraction}{
	name=branching fraction,
	description={The branching fraction of a decay denotes how often this decay occurs relative to the total number of possible decays ({\em absolute} branching fraction) or relative to the occurrences of another specific decay mode ({\em relative} branching fraction)}}

\newglossaryentry{impact parameter}{
	name=impact parameter,
	description={The impact parameter is the three-dimensional point of closest approach (\gls{PCA}) of the reconstructed track with respect to the {\em real} starting point. It is a quality parameter of a single reconstructed track and has to be determined by simulations}}

\newglossaryentry{resonance}{
	name=resonance,
	description={A resonance is the peak in reaction probability located around a certain energy. These peaks are associated with subatomic particles and their excitations. The width $\Gamma$ of a resonance is related to the life time of the corresponding particle (or its excited state)}}

\newglossaryentry{cross section}{
	name=cross section,
	description={The cross section is a hypothetical area which describes the likelihood of a reaction in a particle collision. It is different from the geometrical cross sections of the involved particles, and depends strongly on the collision energy. The typical unit is ``barn'', where $1\,\mathrm{b} = 10^{-28}\,\mathrm{m}^2$ ($100\,\mathrm{fm}^2$)}}
	
\newglossaryentry{luminosity}{
	name=luminosity,
	description={The luminosity $\mathcal{L}$ is one of the main parameters of a particle colliders, as it translates directly to the reaction rate $R$ via $R = \sigma \mathcal{L}$, where $\sigma$ is the \gls{cross section} of the reaction in question. The luminosity can be tuned by proper adjustment of the collider}}
	
\newglossaryentry{bunch}{
	name=bunch,
	description={A bunch is a substructure of the particle beam. The particles are grouped to form ``clouds'' of several million particles travelling together},
	plural=bunches}
	
\newglossaryentry{pion}{
	name=pion,
	description={A pion (symbol: $\uppi$) is any of a group of three \glspl{meson}. In the quark model they are understood to be bound states of any combination of up \glspl{quark} and down \glspl{quark} and their \anti particles}}

\newglossaryentry{sms}{
	name=silicon microstrip sensor,
	description={A silicon microstrip sensor is in principle a large-area \gls{bulk} diode manufactured on a silicon \gls{wafer}, with one ore both electrodes segmented to form strips. When an ionising particle traverses the sensor, the segments near the particle will register a signal, thus measuring the position of the particle}}
	
\newglossaryentry{bulk}{
	name=bulk,
	description={The bulk is the volume of a silicon \gls{substrate}, as opposed to the surface of the \gls{substrate}}}

\newglossaryentry{wafer}{
	name=wafer,
	description={A wafer is a thin, approximately round slice of pure silicon, which is used to produce microchips or -- as in our case -- sensors}}

\newglossaryentry{drift chamber}{
	name=drift chamber,
	description={In a drift chamber many parallel wires (sense wires) are arranged as a grid and put on high voltage, with the metal casing or field forming wires being on ground potential. The chamber is filled with gas and a traversing particle leaves a trace of ions and \glspl{electron}, which drift toward the nearest sense or field wires, respectively. By labelling the wires with a current signal it is possible to measure the track of the traversing particle. In a drift chamber, the timing of the pulse and thus the drift time of the charges is measured as well, which improves the spatial resolution within the cell}}
	
\newglossaryentry{Cherenkow counter}{
	name=Cherenkov counter,
	description={A Cherenkov counter uses the Cherenkov effect to distinguish particle types by determining their velocity. The speed of light in a material is lower than the speed of light in vacuum. A particle traversing a material at a velocity higher than the velocity of light {\em in that material} causes the material to radiate \glspl{photon}. The angle of this radiation relative to the direction of the particle depends on the velocity of the particle. Together with an energy measurement this allows to determine the mass  and thus the identity of the particle}}

\newglossaryentry{scintillator}{
	name=scintillator,
	description={A scintillator is a material that exhibits scintillation -- the property of luminescence --  when excited by ionising radiation. Luminescent materials, when struck by an incoming particle, absorb its energy and scintillate, i.e.\ re-emit the absorbed energy in form of light}}

\newglossaryentry{calorimeter}{
	name=calorimeter,
	description={A calorimeter is an experimental apparatus that measures the energy of particles. Most particles enter the calorimeter and initiate a particle shower and the particles' energy is deposited in the calorimeter, collected, and measured. The energy may be measured in its entirety, requiring total containment of the particle shower, or it may be sampled. Typically, calorimeters are segmented transversely to provide information about the direction of the particle or particles, as well as the energy deposited, and longitudinal segmentation can provide information about the identity of the particle based on the shape of the shower as it develops}}
	
\newglossaryentry{occupancy}{
	name=occupancy,
	description={The occupancy is defined as the fraction of responding (hit) readout channels per triggered event. A high occupancy ($\gg 10\%$) renders the discrimination of individual particle tracks impossible due to an exploding number of combinatorial candidates}}

\newglossaryentry{photomultiplier}{
	name=photomultiplier,
	description={Photomultipliers are extremely sensitive detectors of light in the ultraviolet, visible, and near-infrared ranges of the electromagnetic spectrum. These detectors multiply the current produced by incident light by as much as 100 million times in multiple stages, enabling individual photons to be detected when the incident flux of light is very low}}

\newglossaryentry{granularity}{
	name=granularity,
	description={Granularity is the extent to which a system is broken down into small parts. In terms of particle detectors it means geometrical segmentation for position measurements}}
	
\newglossaryentry{endcap}{
	name=endcap,
	description={Usual particle detectors at colliders are shaped like a cylinder around the beam axis. In contrast to the \gls{barrel} part, the endcap is the circular part of this detector, which closes this cylinder on both sides, like lids}}

\newglossaryentry{barrel}{
	name=barrel,
	description={Usual particle detectors at colliders are shaped like a cylinder around the beam axis. In contrast to the \gls{endcap} part, the barrel part is the cylindrical part of this detector}}

\newglossaryentry{region of interest}{
	name=region of interest,
	description={The region of interest is the region inside the \gls{PXD} which is flagged for readout when an event is triggered. Due to the long integration time and high channel count of the sensors it is not feasible to read out every pixel. The regions of interest are found by fast online \gls{tracking} using the \gls{SVD} data},
	plural=regions of interest}

\newglossaryentry{forward}{
	name=forward,
	description={The forward part of \gls{BelleII} is the side of positive $z$ direction. See section~\ref{sec:acceptance} for details}}

\newglossaryentry{backward}{
	name=backward,
	description={The backward part of \gls{BelleII} is the side of negative $z$ direction. See section~\ref{sec:acceptance} for details}}
	
\newglossaryentry{acceptance}{
	name=acceptance,
	description={The acceptance region of \gls{BelleII} is the polar angle region $17\degree \leq \theta \leq 150\degree$. This region is equipped with particle-sensitive devices. See section~\ref{sec:acceptance} for details}}

\newglossaryentry{radiation length}{
	name=radiation length,
	description={The radiation length is the mean distance over which a high-energy electron loses all but 1/e of its energy by \gls{bremsstrahlung}, where e is Euler's number}}

\newglossaryentry{hermetic}{
	name=hermetic,
	description={In common language ``hermetic'' means ``airtight''. Applied to particle detectors, it means the absence of blind spots in the detector, which are not equipped with particle sensing devices. Thus, a hermetic detector can measure a particle regardless of its direction (as long as it is inside the \gls{acceptance} region)}}

\newglossaryentry{ladder}{
	name=ladder,
	description={The ladder is the basic building block of the \gls{SVD}. It combines the sensors, the front-end readout electronics and the mechanical support structure to an indivisible unit, which is used to build a detector \gls{layer}}}

\newglossaryentry{layer}{
	name=layer,
	description={A detector layer is an approximately cylindrical surface of sensing devices, usually in the \gls{barrel} part of the detector. A layer of the \gls{SVD} is composed of \glspl{ladder}}}
	
\newglossaryentry{multiple scattering}{
	name=multiple scattering,
	description={Multiple scattering is the deflection of a particle when it traverses a material, where information about the origin of the particle is obscured. For \gls{tracking} purposes it is therefore advisable to minimise the \gls{material budget} causing the scattering}}

\newglossaryentry{material budget}{
	name=material budget,
	description={In contrast to the physical thickness of a scatterer, the material budget of a scatterer is the thickness in units of the \gls{radiation length}, usually expressed in \%. This material budget has to be minimised in order to minimise the effects of \gls{multiple scattering}. The material budget can be minimised by reducing the physical thickness of a scatterer, and/or by using a material with a large \gls{radiation length}}}

\newglossaryentry{tracking}{
	name=tracking,
	description={Tracking (also called track reconstruction) is the process of combining individual three-dimensional space points measured by several \glspl{layer} of track sensitive devices, to reconstruct the trajectory of the physical particle. Efficient tracking algorithms allow to extract a maximum of information from the track, such as the curvature in the magnetic field for measuring the momentum of the particle}}
	
\newglossaryentry{rib}{
	name=rib,
	description={The rib is the main support structure of a \gls{ladder} of the \gls{BelleII} \gls{SVD}. It is a sandwich of two carbon fiber plies separated by an Airex\TReg{} core. Each \gls{ladder} is supported by two ribs}}
	
\newglossaryentry{area moment of inertia}{
	name=area moment of inertia,
	description={The area moment of inertia, also known as second moment of area, is a geometrical property of an area which reflects how its points are distributed with regard to an arbitrary axis. It is closely linked to the calculation of stiffness and straws of bent beams}}
	
\newglossaryentry{wire bonding}{
	name=wire bonding,
	description={Wire bonding is the primary method of making interconnections between an integrated circuit (\gls{IC}) and a device package, or (occasionally), directly onto a printed circuit board (\gls{PCB}) during semiconductor device fabrication. Although less common, wire bonding can be used to connect an \gls{IC} to other electronics or to connect from one \gls{PCB} to another. Wire bonding is generally considered the most cost-effective and flexible interconnect technology, and is used to assemble the vast majority of semiconductor packages. In the assembly of \gls{sms} modules, wire bonding is used to connect the strips to the \gls{pitch adapter}, and further to the readout chip}}
	
\newglossaryentry{end-ring}{
	name=end-ring,
	description={The end-rings form a mechanical support structure of the \gls{BelleII} \gls{SVD}, onto which the \glspl{ladder} are mounted. They arrange the \glspl{ladder} in a roughly cylindrical shape}}

\newglossaryentry{acceptor}{
	name=acceptor,
	description={In semiconductor physics, an acceptor is a \gls{dopant} atom that can form a \gls{p-type} region. For example, when silicon (Si), having four valence electrons, needs to be doped as a \gls{p-type} semiconductor, elements from group III like boron (B) or aluminium (Al), having three valence electrons, can be used. When substituting a Si atom in the crystal lattice, the three valence electrons of boron form covalent bonds with three of the Si neighbours but the bond with the fourth neighbour remains unsatisfied. At room temperature, an electron from the neighbouring bond will jump to repair the unsatisfied bond thus leaving a hole (a place where an electron is deficient) in a chain-like process, which results in the hole moving around the crystal and able to carry a current thus acting as a charge carrier. The initially neutral acceptor becomes negatively charged (ionised)}}

\newglossaryentry{donor}{
	name=donor,
	description={In semiconductor physics, a donor is a \gls{dopant} atom that can form an \gls{n-type} region. For example, when silicon (Si), having four valence electrons, needs to be doped as an \type{n} semiconductor, elements from group V like phosphorus (P) or arsenic (As) can be used because they have five valence electrons. When substituting a Si atom in the crystal lattice, four of the valence electrons of phosphorus form covalent bonds with the neighbouring Si atoms but the fifth one remains weakly bonded. At room temperature, all the fifth electrons are liberated, can move around the Si crystal and can carry a current and thus act as charge carriers. The initially neutral donor becomes positively charged (ionised)}}

\newglossaryentry{p-type}{
	name=p-type,
	description={Semiconductors doped with \gls{acceptor} impurities are called \type{p}, and have an excess of \glspl{hole} as majority charge carriers}}

\newglossaryentry{n-type}{
	name=n-type,
	description={Semiconductors doped with \gls{donor} impurities are called \type{n} and have an excess of \glspl{electron} as majority charge carriers}}
	
\newglossaryentry{hole}{
	name=hole,
	description={In semiconductor physics, a hole (or defect electron) is a positive charge carrier, which is understood to be a missing electron in an otherwise fully occupied valence band}}
	
\newglossaryentry{p-side}{
	name=p-side,
	description={The \side{p} of a \glslink{DSSD}{double-sided silicon sensor} is the one with \gls{p-type} implanted strips}}
	
\newglossaryentry{n-side}{
	name=n-side,
	description={The \side{n} of a \glslink{DSSD}{double-sided silicon sensor} is the one with \gls{n-type} implanted strips}}

\newglossaryentry{junction side}{
	name=junction side,
	description={The junction side of a \glslink{DSSD}{double-sided silicon sensor} is the one where the \pnjunction{} is located. This is the \gls{p-side} for \gls{n-type} material, and the \gls{n-side} for \gls{p-type} material}}

\newglossaryentry{ohmic side}{
	name=ohmic side,
	description={The ohmic side of a \glslink{DSSD}{double-sided silicon sensor} is the one where \emph{no} \pnjunction{} is located. On the ohmic side, a highly doped region is connected to the lowly doped \gls{bulk}, forming an ohmic contact.This is the \gls{n-side} for \gls{n-type} material, and the \gls{p-side} for \gls{p-type} material}}
	
\newglossaryentry{AC-coupled}{
	name=AC-coupled,
	description={An \gls{AC}-coupled sensor strip is connected to the readout electronics via a coupling capacitor. Being only conductive for \gls{AC} currents, it filters the strip's \gls{dark current} (a \gls{DC} current) away. The capacitance of this coupling is called ``\gls{cac}''}}

\newglossaryentry{dark current}{
	name=dark current,
	description={A sensor's dark current is the current through the sensor when there is \emph{no} signal present. Since a sensor can be thought of as a large \gls{bulk} \gls{diode}, the dark current is basically the \gls{leakage current} of this \gls{diode}}}

\newglossaryentry{bias resistor}{
	name=bias resistor,
	description={The bias resistor connects the strips of a \gls{sms} to the \gls{bias ring}, which is further connected to the power supply. It is made of \gls{polysilicon}}}

\newglossaryentry{intermediate strip}{
	name=intermediate strip,
	description={An intermediate strip is a strip which is implanted, but has no metallisation to be read out. Its purpose is to pick up the part of the signal generated between the actual readout strips and couple it to them capacitively, thus improving the \gls{charge sharing} between the readout strips without doubling the number of readout channels}}

\newglossaryentry{charge sharing}{
	name=charge sharing,
	description={In \glspl{sms}, charge sharing is the beneficial effect where the charge generated by a traversing particle is collected by two or more adjacent strips. This makes the position measurement more accurate, as long as the amount of the collected charge is measured. Charge sharing is not to be confused with crosstalk, where the signal picked up by a strip is capacitively coupled to an adjacent strip. Crosstalk does \emph{not} improve the position measurement}}

\newglossaryentry{pitch}{
	name=pitch,
	description={Strip distance from center to center. Distinguish between the pitch of all implanted strips and the pitch of the readout strips (in case of a sensor with \glspl{intermediate strip}). Usually -- unless otherwise noted -- pitch refers to the distance between the readout strips},
	plural=pitches}

\newglossaryentry{active area}{
	name=active area,
	description={The active area of a \gls{sms} is the area where it is sensitive to particles, i.e.\ the area where the electrode is segmented to form strips. To simplify things, in this thesis the active area is defined as the area outlined by the \gls{bias ring}}}

\newglossaryentry{bias ring}{
	name=bias ring,
	description={The bias ring encloses the \gls{active area} and is connected to the power supply for biasing the sensor. It hands the potential over to the strips via the \glspl{bias resistor}, \gls{FOXFET} or \gls{punch through} biasing}}
	
\newglossaryentry{pad}{
	name=pad,
	description={A pad is a small metalized area designated for contacting the sensor with probe needles or wire bonds through a window in the \gls{passivation}}}

\newglossaryentry{passivation}{
	name=passivation,
	description={The passivation is the final protection layer of \gls{CVD} deposited silicon dioxide (or similar materials depending on the vendor). This layer renders the sensor insensitive to a range of environmental influences, such as oxidation and corrosion}}

\newglossaryentry{depletion}{
	name=depletion,
	description={In semiconductor physics, the depletion region, also called \gls{scr}, is an insulating region within a conductive, doped semiconductor material where the mobile charge carriers have diffused away, or have been forced away by an electric field. The only elements left in the depletion region are ionised \gls{donor} or \gls{acceptor} impurities}}

\newglossaryentry{polysilicon}{
	name=polysilicon,
	description={Polycrystalline silicon, also called polysilicon, is a material consisting of small silicon crystals. It differs from single-crystal silicon -- used for electronics and sensors -- and from amorphous silicon -- used for thin film devices}}
	
\newglossaryentry{guard ring}{
	name=guard ring,
	description={The guard ring is a structure for high voltage protection of a \gls{sms}. It is on floating potential between the \gls{bias ring} and the edge of the sensor, where it creates a controlled, homogeneous voltage drop}}
	
\newglossaryentry{p-stop}{
	name=p-stop,
	description={The \pstop{} blocking method is a method to ensure electrical strip separation on the \gls{n-side} of \glspl{sms}. It features \gls{p-type} \glspl{implantation} between the \gls{n-type} strips. These \glspl{implantation} are called \pstop{}s. The p-stop implants are usually on floating potential, only for special measurements one might want to put them on a defined potential}}
	
\newglossaryentry{atoll}{
	name=atoll,
	description={A variant of the \gls{p-stop} blocking method. Each \gls{n-type} strip is surrounded by its individual ``ring-like'' \gls{p-type} \gls{implantation}}}

\newglossaryentry{interstrip resistance}{
	name=interstrip resistance,
	description={The interstrip resistance is the ohmic resistance between an implanted strip and its immediate neighbour, disregarding the \glspl{bias resistor}}}

\newglossaryentry{VA1TA}{
	name=VA1TA,
	description={The VA1TA is the chip of the former \gls{Belle} experiment used for reading out the \glspl{DSSD} of the \gls{SVD2}}}

\newglossaryentry{SVD2}{
	name=SVD2,
	description={The SVD2 is the upgraded Silicon Vertex Detector of the former \gls{Belle} experiment. Not to be mixed up with the \gls{BelleII} \gls{SVD}, which is being built at the moment}}

\newglossaryentry{apv}{
	name=APV25,
	description={The future \gls{BelleII} experiment will use the \apv{} chip for reading out the \glspl{DSSD} of the \gls{SVD}. It was developed and is used for the \gls{CMS} experiment at the \gls{LHC} at \gls{CERN}. The acronym stands for Analog Pipeline Voltage (0.25um)}}

\newglossaryentry{radio frequency}{
	name=radio frequency,
	description={The radio frequency is the frequency at which the accelerator stage of a collider operates. It divides the particles into \glspl{bunch} at a distance of the radio frequency's wave length, and is therefore closely related to the \gls{bunch} collision rate. However, only every other \gls{bunch} is filled with particles, so that the collision rate is only half the nominal value; collisions occur approximately every 4\ns.}}
	
\newglossaryentry{origami}{
	name=Origami,
	description={In the \origami{} chip-on-sensor concept, the \gls{apv} readout chips are mounted on a flexible \gls{hybrid} board directly on the sensor, while the strips of the other side are connected using flexible \glspl{pitch adapter} wrapped around the edge of the sensor (hence the name \origami{}). This puts the readout chip close to the strips -- minimising the capacitive load and thus the noise -- while maintaining an acceptable \gls{material budget}}}

\newglossaryentry{pitch adapter}{
	name=pitch adapter,
	description={A pitch adapter is a flexible or rigid structure implementing conductive traces for connecting the sensor strips to the readout chip. On one side, the \gls{pitch} of these connections matches the \gls{pitch} of the sensor strips, on the other side it matches the \gls{pitch} of the readout chip's input channels, which is usually narrower than the sensor side. Pitch adapters are also called ``fanouts''}}

\newglossaryentry{snr}{
	name=signal-to-noise ratio,
	description={Signal-to-noise ratio (often abbreviated \gls{SNR} or S/N) is a measure used in science and engineering that compares the level of a desired signal to the level of background noise. It is defined as the ratio of signal power to the noise power, often expressed in decibels. A ratio higher than 1:1 (greater than 0\,dB) indicates more signal than noise. While \gls{SNR} is commonly quoted for electrical signals, it can be applied to any form of signal. In terms of \glspl{sms}, the signal-to-noise ratio is the division of the signal released by a traversing particle and the quadratically added noise of all hit strips. A value in the order of 10 or higher is needed to safely detect ionising particles and perform \gls{tracking}}}

\newglossaryentry{pedestal subtraction}{
	name=pedestal subtraction,
	description={The pedestal subtraction is a correction algorithm in the data acquisition, which ensures a mean readout channel reading of zero when no signal is present}}

\newglossaryentry{cm}{
	name=common mode correction,
	description={The common mode correction is a correction algorithm in the data acquisition which removes correlated noise of applying to all strips of a \gls{sms}}}
	
\newglossaryentry{zero suppression}{
	name=zero suppression,
	description={The zero suppression removes data of strips with no signal -- which is usually the vast majority -- and thus reduces the data rate drastically}}
	
\newglossaryentry{htf}{
	name=hit time finding,
	description={The hit time finding is an algorithm in the data acquisition of the \gls{BelleII} \gls{SVD} which reconstructs the exact timing of the signal's maximum}}

\newglossaryentry{latent heat}{
	name=latent heat,
	description={Latent heat is the heat released or absorbed by a body or a thermodynamic system during a process that occurs without a change in temperature. A typical example is a change of state of matter, meaning a phase transition such as the evaporation of \gls{CO2}}}

\newglossaryentry{conduction band}{
	name=conduction band,
	description={The conduction band is the range of \gls{electron} energies sufficient to free an \gls{electron} from binding with its atom to move freely within the atomic lattice of the material as a delocalized \gls{electron}}}

\newglossaryentry{valence band}{
	name=valence band,
	description={In solids, the valence band is the highest range of \gls{electron} energies in which \glspl{electron} are normally present at absolute zero temperature}}

\newglossaryentry{band gap}{
	name=band gap,
	description={In solid state physics, a band gap, also called an energy gap, is an energy range in a solid where no \gls{electron} states can exist. In graphs of the electronic band structure of solids, the band gap generally refers to the energy difference (in eV) between the top of the \gls{valence band} and the bottom of the \gls{conduction band} in insulators and semiconductors. For metals the \gls{conduction band} and the \gls{valence band} overlap, removing the band gap}}
	
\newglossaryentry{energy band}{
	name=energy band,
	description={In solid-state physics, the electronic band structure (or simply band structure) of a solid describes those ranges of energy that an \gls{electron} within the solid may have (called energy bands, allowed bands, or simply bands), and ranges of energy that it may not have (called \glspl{band gap} or forbidden bands)}}

\newglossaryentry{intrinsic}{
	name=intrinsic,
	description={An intrinsic semiconductor is a pure semiconductor material, not altered by \gls{doping}}}

\newglossaryentry{extrinsic}{
	name=extrinsic,
	description={An extrinsic semiconductor is a semiconductor material whose electrical properties have been altered by \gls{doping}}}

\newglossaryentry{doping}{
	name=doping,
	description={In semiconductor production, doping intentionally introduces impurities into an extremely pure (``\gls{intrinsic}'') semiconductor for the purpose of modulating its electrical properties.
	The impurities are dependent upon the type of semiconductor.
	Lightly and moderately doped semiconductors are referred to as \gls{extrinsic}}}

\newglossaryentry{phonon}{
	name=phonon,
	description={A phonon is a  quasi-particle and understood to be the quantum of vibration modes of the crystal lattice}}
	
\newglossaryentry{kB}{
	name=Boltzmann constant,
	description={The Boltzmann constant \kB, named after Ludwig Boltzmann, is a physical constant relating energy at the individual particle level with temperature. It is the gas constant $R$ divided by the Avogadro constant $N_\mathrm{A}$: $\kB = \nicefrac{R}{N_\mathrm{A}} = 1.3806 \times 10^{-23} \nicefrac{\mathrm{J}}{\mathrm{K}} = 8.6173 \times 10^{-5} \nicefrac{\eV}{\mathrm{K}}$}}

\newglossaryentry{EF}{
	name=Fermi energy,
	description={At the Fermi energy the \gls{FDpdf} has a value of 1/2},
	plural=Fermi energies}
	
\newglossaryentry{FDpdf}{
	name=Fermi-Dirac probability density function,
	description={In quantum statistics, Fermi-Dirac (F-D) statistics describes distribution of particles in a system comprising many identical particles that obey the Pauli exclusion principle. It applies to identical particles with half-odd-integer spin in a system in thermal equilibrium. It is most commonly applied to \glspl{electron}, which are \glspl{fermion} with spin 1/2. The Fermi-Dirac \gls{pdf} reads: $n_i = \frac{1}{e^{(\epsilon_i-\mu)/\kB T}+1}$, where \kB\ is Boltzmann's constant, $T$ is the absolute temperature, $\epsilon_i$ is the energy of the single-particle state , and $\mu$ is the total chemical potential. For the case of \glspl{electron} in a semiconductor, $\mu$ is typically called the Fermi level or electrochemical potential}}

\newglossaryentry{dispersion relation}{
	name=dispersion relation,
	description={In physics and electrical engineering, dispersion relations describe the effect of dispersion from a medium on the properties of a wave traveling within that medium. A dispersion relation connects different properties of the wave such as its energy, frequency, wavelength and wavenumber. From these relations, the phase velocity and group velocity of the wave have convenient expressions which can thereby determine a refractive index of the medium}}
	
\newglossaryentry{drift}{
	name=drift,
	description={Drift -- as opposed to \gls{diffusion} -- is the motion of charge carriers caused by an electric field $\vec{E}$. The charge carriers will move along the electric field lines, where the direction of this movement is determined by the charge of the carrier}}

\newglossaryentry{diffusion}{
	name=diffusion,
	description={Diffusion -- as opposed to \gls{drift} -- is the random motion of charge carriers which tends to level out geometrical charge carrier distribution inhomogeneities}}

\newglossaryentry{pnjunction}{
	name=p-n-junction,
	description={A p-n junction is a boundary or interface between two types of semiconductor material, \gls{p-type} and \gls{n-type}, inside a single crystal of semiconductor. It is created by \gls{doping}, for example by ion \gls{implantation}, \gls{diffusion} of \glspl{dopant}, or by epitaxy (growing a layer of crystal doped with one type of dopant on top of a layer of crystal doped with another type of dopant)}}

\newglossaryentry{bias voltage}{
	name=bias voltage,
	description={Biasing in electronics is the method of establishing predetermined voltages or currents at various points of an electronic circuit for the purpose of establishing proper operating conditions in electronic components. Many electronic devices whose function is to process time-depending (\gls{AC}) signals also require a constant (\gls{DC}) current or voltage to operate correctly. In terms of \glspl{sms}, the bias voltage is the voltage applied between the two side of the sensor, which depletes the sensor and establishes the \gls{drift} field for the signal charge carriers}}

\newglossaryentry{ehpair}{
	name=\mbox{e-h-pair},
	description={The electron-hole pair is the fundamental unit of charge carrier generation and recombination, corresponding to an \gls{electron} transitioning between the \gls{valence band} and the \gls{conduction band}}}

\newglossaryentry{bremsstrahlung}{
	name=bremsstrahlung,
	description={Bremsstrahlung (from German bremsen ``to brake'' and Strahlung ``radiation'', i.e.\ ``braking radiation'' or ``deceleration radiation'') is electromagnetic radiation produced by the deceleration of a charged particle when deflected by another charged particle, typically an \gls{electron} by an atomic nucleus. The moving particle loses kinetic energy, which is converted into a \gls{photon} because energy is conserved. The term is also used to refer to the process of producing the radiation}}

\newglossaryentry{crystal momentum}{
	name=crystal momentum,
	description={In solid-state physics, crystal momentum or quasimomentum is a \\ \mbox{momentum-like} vector associated with \glspl{electron} in a crystal lattice. It is defined by the associated wave vectors $k$ of this lattice, according to $p_\mathrm{crystal} = \hbar k$, where $\hbar = 6.626 \times 10^{-34}\,\mathrm{Js} = 4.1357 \times 10^{-15}\eV\mathrm{s}$ is the reduced Planck's constant. Often the wave number $k$ is wrongly called crystal momentum},
	plural=crystal momenta}

\newglossaryentry{ionisation}{
	name=ionisation,
	description={Ionisation is the process of converting an atom or molecule into an ion by adding or removing charged particles such as \glspl{electron} or ions}}
	
\newglossaryentry{mollersc}{
	name=M\o ller scattering,
	description={M\o ller scattering is the name given to \gls{electron}-\gls{electron} scattering in Quantum Field Theory, named after the danish physicist Christian M\o ller. The \gls{electron} interaction that is idealised in M\o ller scattering forms the theoretical basis of many familiar phenomena such as the repulsion of \glspl{electron} in the Helium nucleus}}
	
\newglossaryentry{bhabasc}{
	name=Bhabha scattering,
	description={In quantum electrodynamics, Bhabha scattering is the \gls{electron}-\gls{positron} interaction process, where the particle either scatter or annihilate and create a new \gls{electron}-\gls{positron} pair}}
	
\newglossaryentry{mip}{
	name=minimum ionising particle,
	description={In physics, a minimum ionising particle (or \gls{MIP}) is a particle whose mean energy loss rate through matter is at (or close to) the possible minimum}}
	
\newglossaryentry{BBeq}{
	name=Bethe-Bloch equation,
	description={In nuclear physics and theoretical physics, charged particles moving through matter interact with the electrons of atoms in the material. The interaction excites or ionises the atoms. This leads to an energy loss of the traversing particle. The Bethe-Bloch formula describes the energy loss per distance travelled of swift charged particles (\glspl{proton}, alpha particles, atomic ions, but not \glspl{electron}) traversing matter (or alternatively the stopping power of the material)}}

\newglossaryentry{Landau_pdf}{
	name=Landau probability density function,
	description={In probability theory, the Landau \gls{pdf} is a probability distribution named after Lev Landau. Because of the function's long tail, the moments of the distribution, like mean or variance, are undefined. $p(x) = \frac{1}{2 \pi i} \int_{c-i\infty}^{c+i\infty}\! e^{s \log s + x s}\,\mathrm{d}s$}}

\newglossaryentry{direct}{
	name=direct,
	description={In semiconductor physics, the \gls{band gap} of a semiconductor is always one of two types, a direct \gls{band gap} or an indirect \gls{band gap}. The \gls{band gap} is called ``direct'' if the momentum of \glspl{electron} and \glspl{hole} is the same in both the \gls{conduction band} and the \gls{valence band}; an \gls{electron} can directly emit a \gls{photon}. In an ``indirect'' gap, a \gls{photon} cannot be emitted because the \gls{electron} must pass through an intermediate state and transfer momentum to the crystal lattice}}
	
\newglossaryentry{minority}{
	name=minority,
	description={Minority charge carriers are the charge carriers with opposite charge compared to the \gls{majority} charge carriers, which exist due to thermal excitation at a much lower concentration}}

\newglossaryentry{majority}{
	name=majority,
	description={Majority charge carriers are the charge carriers introduced by \gls{doping} in an \gls{extrinsic} semiconductor. A \gls{p-type} semiconductor has \glspl{hole} as majority charge carriers, whereas for an \gls{n-type} semiconductor they are \glspl{electron}}}

\newglossaryentry{ccgen}{
	name=charge carrier generation,
	description={In the solid-state physics of semiconductors, carrier generation is a processes by which mobile charge carriers (\glspl{electron} and \glspl{hole}) are created. This can happen by thermal excitation, optical excitation by an absorbed \gls{photon}, and by  charged particles.
The \gls{ehpair} is the fundamental unit of \glslink{ccgen}{generation} and \glslink{ccrec}{recombination}, corresponding to an \gls{electron} transitioning between the \gls{valence band} and the \gls{conduction band}}}
	
\newglossaryentry{ccrec}{
	name=charge carrier recombination,
	description={In the solid-state physics of semiconductors, carrier recombination is a processes by which mobile charge carriers (\glspl{electron} and \glspl{hole}) are eliminated. It is the reverse process of \gls{ccgen}}}	
	
\newglossaryentry{trapping}{
	name=trapping,
	description={Trapping is the process of interrupting a charge carrier's \gls{drift} or \gls{diffusion} by binding it to a lattice impurity. The impurity is often charged in that process, representing a scattering center further on}}

\newglossaryentry{scr}{
	name=space charge region,
	description={In semiconductor physics, the \gls{depletion} region, also called depletion layer, depletion zone, junction region or the space charge region, is an insulating region within a conductive, doped semiconductor material where the mobile charge carriers have diffused away, or have been forced away by an electric field. The only elements left in the depletion region are ionised \gls{donor} or \gls{acceptor} impurities}}		

\newglossaryentry{Vbi}{
	name=built-in voltage,
	description={The built-in voltage is the potential difference introduced by the \gls{scr} in a semiconductor's \gls{pnjunction}}}		
	
\newglossaryentry{forward bias}{
	name=forward bias,
	description={In forward bias mode, a \gls{diode} becomes conductive. \Glspl{electron} are injected into the \gls{p-type} material and \glspl{hole} into the \gls{n-type} material}}		

\newglossaryentry{reverse bias}{
	name=reverse bias,
	description={In reverse bias mode, a \gls{diode} becomes an insulator. The \gls{p-type} \gls{bulk} band edges are raised relative to the \gls{n-type} \gls{bulk} by the reverse bias voltage, so that the two \gls{bulk} occupancy levels are separated again by an energy determined by the applied voltage}}		

\newglossaryentry{resistivity}{
	name=resistivity,
	description={Electrical resistivity (also known as resistivity, specific electrical resistance, or volume resistivity) quantifies how strongly a given material opposes the flow of electric current}}		

\newglossaryentry{diode}{
	name=diode,
	description={In electronics, a diode is a two-terminal component with asymmetric conductance, it has low (ideally zero) resistance to current flow in one direction, and high (ideally infinite) resistance in the other. A semiconductor diode, the most common type today, is a crystalline piece of semiconductor material with a \gls{pnjunction} connected to two electrical terminals}}		

\newglossaryentry{leakage current}{
	name=leakage current,
	description={In semiconductor devices, leakage is a quantum phenomenon where mobile charge carriers (\glspl{electron} or \glspl{hole}) tunnel through an insulating region. Leakage increases exponentially as the thickness of the insulating region decreases. Tunneling leakage can also occur across semiconductor junctions between heavily doped \gls{p-type} and \gls{n-type} semiconductors. Other than tunneling via the gate insulator or junctions, carriers can also leak between source and drain terminals of a \gls{MOS} transistor. See also \gls{dark current}}}		

\newglossaryentry{IV-curve}{
	name=IV-curve,
	description={The IV-curve shows the \gls{leakage current} of a \gls{diode} or a \gls{sms} as function of the \gls{bias voltage}. The current at the \gls{VFD} is a quality indicator for a \gls{sms}}}		

\newglossaryentry{CV-curve}{
	name=CV-curve,
	description={The CV-curve shows the total capacitance of a \gls{diode} or a \gls{sms} as function of the \gls{bias voltage}. It can be used to determine the \gls{VFD}}}		

\newglossaryentry{VFD}{
	name=full depletion voltage,
	description={When ramping up a reverse \gls{bias voltage} to a sensor, the \gls{depletion} region at the \pnjunction{} is increased, until it reaches the other side of the sensor and the whole \gls{bulk} to be free of charge carriers. The voltage needed to fully deplete the sensor is called ``full depletion voltage''. It is mainly determined by the \gls{resistivity} of the \gls{wafer}. It is furthermore an important benchmark for the operation of a \gls{sms}, as the operating voltage should always be larger than the full depletion voltage to ensure a maximum of detection efficiency}}

\newglossaryentry{SC}{
	name=Schottky contact,
	description={A Schottky contact (or Schottky barrier), named after Walter H.\ Schottky, is a potential energy barrier for \glspl{electron} formed at a metal-semiconductor junction. Schottky barriers have rectifying characteristics, suitable for use as a diode}}		

\newglossaryentry{electron affinity}{
	name=electron affinity,
	description={In chemistry and atomic physics, the electron affinity of an atom or molecule is defined as the amount of energy released when an \gls{electron} is added to a neutral atom or molecule to form a negative ion}}		

\newglossaryentry{degenerate}{
	name=degenerate,
	description={A degenerate semiconductor is a semiconductor with such a high level of doping that the material starts to act more like a metal than a semiconductor}}		

\newglossaryentry{tunneling}{
	name=tunneling,
	description={Quantum tunnelling refers to the quantum mechanical phenomenon where a particle tunnels through a barrier that it classically could not surmount. This plays an essential role in several physical phenomena, such as the nuclear fusion that occurs in main sequence stars like the Sun. It has important applications to modern devices such as the tunnel \gls{diode} and the scanning tunneling microscope}}		

\newglossaryentry{inversion}{
	name=inversion,
	description={Inversion situation occurs in a \gls{pnjunction}, when a considerable amount of \gls{minority} charge carriers are aggregated at an interface between a semiconductor and another material. This situation can occur for strongly bent \glspl{energy band}. In case of \gls{p-type} material, the formed \gls{inversion} layer consists of \glspl{electron} and shorts \gls{n-type} doped regions at the surface. It therefore has to be interrupted by a \gls{p-stop} or \gls{p-spray} \gls{implantation}}}		

\newglossaryentry{accumulation}{
	name=accumulation,
	description={Accumulation situation occurs in a \gls{pnjunction}, when a considerable amount of \gls{majority} charge carriers are accumulated at an interface between a semiconductor and another material. This situation can occur for weakly bent \glspl{energy band}, e.g.\ due to the presence of fixed positive oxide charges. In case of \gls{n-type} material, the formed accumulation layer consists of \glspl{electron} and shorts \gls{n-type} doped regions at the surface. It therefore has to be interrupted by a \gls{p-stop} or \gls{p-spray} \gls{implantation}}}		

\newglossaryentry{flat-band}{
	name=flat-band,
	description={A \gls{MOS} structure is in flat-band situation, when the intrinsic \gls{energy band} deformation (e.g.\ by different \glspl{work function}) is compensated by an external voltage so that the bands are flat. The external voltage is then called ``flat-band voltage''. The lower this voltage, the lower is the concentration of fixed oxide charges, which is a quality indicator for \glspl{sms}}}		

\newglossaryentry{mpv}{
	name=most probable value,
	description={In probability calculation, the most probable value is the peak of a \gls{pdf}. In general, it is different from the mean value (expectation value), and it can be used for distributions lacking defined moments, like the \gls{Landau_pdf}}}		

\newglossaryentry{cce}{
	name=charge collection efficiency,
	description={The charge collection efficiency is a quality indicator for \glspl{sms}. Ideally, every single charge carrier released by a traversing particle should drift all the way to the electrodes, inducing the maximum signal possible. In reality, lattice imperfections trap and scatter the charge carriers, leading to a reduced charge collection efficiency. Note that the exact amount of ionised charge carriers is not known, so that the charge collection efficiency can only be an estimate}}		

\newglossaryentry{state density}{
	name=state density,
	description={In solid-state and condensed matter physics, the density of states (DOS) of a system describes the number of states per interval of energy at each energy level that are available to be occupied by \glspl{electron}. Unlike isolated systems, like atoms or molecules in gas phase, the density distributions are not discrete like a spectral density but continuous}}		
	
\newglossaryentry{implantation}{
	name=implantation,
	description={Ion implantation is an engineering process by which ions of a material are accelerated in an electrical field and impacted into a solid. This process is used to change the physical, chemical, or electrical properties of the solid. Ion implantation is used in semiconductor device fabrication and in metal finishing, as well as various applications in materials science research}}		

\newglossaryentry{Lorentz angle}{
	name=Lorentz angle,
	description={When charge carriers drift in a combined electrical and magnetically field, they are deflected from the electric field lines. The deflection angle is called ``Lorentz angle''}}		

\newglossaryentry{weighting field}{
	name=weighting field,
	description={According to~\cite{ref:ramo} the weighting field is the electric field which would exist at the charge carrier's instantaneous position under the following circumstances: the charge carrier is removed, the chosen strip is raised to unit potential, and all other strips are grounded. The weighting field allows the calculation of the signal induced in a particular strip by a moving charge carrier as a function of the relative position of strip and charge carrier}}		

\newglossaryentry{pixel sensor}{
	name=pixel sensor,
	description={In particle physics, a pixel sensor is a position sensitive device which delivers unambiguous 2D position measurements of traversing charged particles. In contrast to a \gls{sms}, a silicon pixel sensor segments one electrode in a two-dimensional array of (rectangular) pixels, each a little silicon sensor in it's own right, with a typical size of around $100 \times 100\mum^2$. An electronic silicon chip, one for each tile is attached, using an almost microscopic spot of solder using the so-called ``bump bonding'' technique, which amplifies the signal. Different architectures include monolithic active pixel sensors (MAPS) and depleted p-channel field effect transistor sensors (\gls{DEPFET})}}		

\newglossaryentry{ghost hit}{
	name=ghost hit,
	description={A ghost hit is a fake position measurement occurring for \glspl{DSSD} or two combined single-sided \glspl{sms} measuring particle positions in two dimensions. When the sensor(s) is (are) hit by two or more particles simultaneously, there are more than two combinatorial possibilities for the measured particle position. The non-real ones are called ``ghost hits'', and the true hit position can only be determined by combinatorics using several layers}}		

\newglossaryentry{stereo angle}{
	name=stereo angle,
	description={The stereo angle is the angle between the strip directions of \gls{n-side} and \gls{p-side} strips of a \gls{DSSD}. It is often $90^\circ$, but in some cases it is beneficial to choose a very small stereo angle, which would imply vastly different position resolutions in the two measured dimensions, but would allow to put the readout electronics on one side of the sensor only}}		

\newglossaryentry{backplane}{
	name=backplane,
	description={The backplane of a single-sided \gls{sms} is the sensor side \emph{not} instrumented with strips. A \gls{DSSD} has no backplane in that sense}}		

\newglossaryentry{cluster}{
	name=cluster,
	description={A cluster is composed of two or more adjacent strips with signals above a predefined threshold. In case of the \gls{BelleII} \gls{SVD}, a signal five times as high as the strip noise defines the \gls{seed strip}; for the \glspl{neighbour strip}, signals three times as high as the respective strip noise are required. The total cluster signal has to be at least five times as high as the total cluster noise}}		

\newglossaryentry{cluster width}{
	name=cluster width,
	description={The cluster width (or cluster size) is the number of strips above threshold for a given hit \gls{cluster}}}		

\newglossaryentry{seed strip}{
	name=seed strip,
	description={The seed strip of a \gls{cluster} is the first strip found to have a signal of five times its single strip noise. The cluster finding algorithm first looks for a strip complying with this threshold, and once found, it continues to search for \glspl{neighbour strip} adjacent to the seed strip}}		

\newglossaryentry{neighbour strip}{
	name=neighbour strip,
	description={The neighbour strip of a cluster is a strip near a \gls{seed strip} with a signal of at least three times its single strip noise}}		

\newglossaryentry{standard deviation}{
	name=standard deviation,
	description={In statistics and probability theory, standard deviation (represented by the symbol sigma, $\sigma$) shows how much variation or dispersion can be expected from the average (mean) value}}		

\newglossaryentry{pdf}{
	name=probability density function,
	description={In probability theory, a probability density function (pdf), or density of a continuous random variable, is a function that describes the relative likelihood for this random variable to take on a given value. The probability for the random variable to fall within a particular region is given by the integral of this variable's density over the region. The probability density function is nonnegative everywhere, and its integral over the entire space is equal to one. A (normalised) histogram is a direct measurement of a probability density function}}		

\newglossaryentry{df}{
	name=distribution function,
	description={In probability theory and statistics, the cumulative distribution function (CDF), or just distribution function, describes the probability that a real-valued random variable $X$ with a given probability distribution will be found at a value less than or equal to $x$. In the case of a continuous distribution, it gives the area under the \gls{pdf} from minus infinity to $x$}}		

\newglossaryentry{eta correction}{
	name=eta correction,
	description={The eta correction is an algorithm for linearising the relationship between the particle hit position and the measured charge center-of-gravity in a \gls{sms}. When a particle hits the sensor between two strips, the charge carriers \gls{drift} towards both strips, but the amount of drifting charge carriers is \emph{not} proportional to the distance from the strips. This effect is especially pronounced for particles traversing the sensor perpendicularly, as is the case for \glspl{beam test}, and for high energetic tracks in the real experiment}}		

\newglossaryentry{beam test}{
	name=beam test,
	description={In a beam test, sensors are exposed to a particle beam to determine their response to real particles. The beam is usually extracted from a particle accelerator or storage ring, and can be tuned to the user's requirements in composition and energy}}		

\newglossaryentry{enc}{
	name=equivalent noise charge,
	description={The equivalent noise charge is the number of electrons one would have to collect from a silicon sensor in order to create a signal equivalent to the noise of this sensor}}		

\newglossaryentry{cac}{
	name=coupling capacitance,
	description={The coupling capacitance is the capacitance created by the implanted strip and the aluminium line atop of it. It is crucial for removing the \gls{DC} component of the \gls{dark current}}}		
	
\newglossaryentry{cint}{
	name=interstrip capacitance,
	description={The interstrip capacitance is the capacitance between two adjacent implanted strips. It is responsible for capacitive coupling of strip signals, which can mimic a cluster of higher \glslink{cluster width}{width} than really present}}		

\newglossaryentry{cback}{
	name=backplane capacitance,
	description={The backplane capacitance is the capacitance between an implanted strip and the other side of the sensor. It is usually calculated as the total capacitance (parallel plate capacitor configuration) divided by the number of strips}}		

\newglossaryentry{punch through}{
	name=punch through,
	description={A \gls{scr} expands as the \gls{reverse bias} voltage is increased. For a p-n-p junction (or equivalently a n-p-n junction) punch-through situation is reached when the \gls{scr} of the reverse biased junction touches the forward biased junction. Then, charge carriers from the forward biased junction will be injected into the reversed biased junction through drift and diffusion. This will lead to an exponential rise in the current with respect to the applied voltage. This effect can be used to apply the \gls{bias voltage} to the strips of a \gls{sms}}}		

\newglossaryentry{AC pad}{
	name=AC~pad,
	description={The AC pad is an area in the metal layer of a strip intended to allow contact to the metal strip. It is called ``AC'' because the metal strip only sees the AC component of the \gls{dark current} and signal}}		

\newglossaryentry{DC pad}{
	name=DC~pad,
	description={The DC pad is a contact area which allows to contact the implanted strip. It is connected to the implanted strip through an \gls{oxide window}, and often also connects the end of the \gls{bias resistor} to the implanted strip}}		

\newglossaryentry{edge ring}{
	name=edge ring,
	description={The edge ring is the outermost structure of a \gls{sms}. It contains inscriptions like the strip numbering, and often features an additional implantation for high-voltage stability and for contacting the \gls{bulk} material using an \gls{n-sub pad}}}		

\newglossaryentry{oxide window}{
	name=oxide window,
	description={The oxide window is a hole in the thin \gls{coupling oxide}, which allows the metallisation to connect to the silicon surface. It is often wrongly called ``via'', which in contrast to an oxide window is a connection between different metal layers}}		

\newglossaryentry{via}{
	name=via,
	description={See \gls{oxide window}}}		

\newglossaryentry{dicing}{
	name=dicing,
	description={Dicing is the process by which dies are separated from a semiconductor \gls{wafer} after complete processing of the \gls{wafer}. The dicing process can be accomplished by scribing and breaking, by mechanical sawing (normally with a machine called a dicing saw) or by laser cutting. During dicing, \glspl{wafer} are typically mounted on dicing tape which has a sticky backing that holds the \gls{wafer} on a thin sheet metal frame}}		
	
\newglossaryentry{dopant}{
	name=dopant,
	description={A dopant, also called a doping agent, is a trace impurity element that is inserted into a substance (in very low concentrations) in order to alter the electrical properties or the optical properties of the substance. In the case of crystalline substances, the atoms of the dopant take the place of elements that were in the crystal lattice of the material. These materials are very commonly either crystals of a semiconductor (silicon, germanium, etc.), for use in solid-state electronics; or else transparent crystals that are used to make lasers of various types}}		

\newglossaryentry{ingot}{
	name=ingot,
	description={An ingot (also called ``boule'') is a material, usually metal, that is cast into a shape suitable for further processing. Non-metallic and semiconductor materials prepared in \gls{bulk} form may also be referred to as ingots, particularly when cast by mold based methods. Silicon ingots are usually cylindrical rods, which are cut into circular \glspl{wafer}}}		

\newglossaryentry{oxidation}{
	name=oxidation,
	description={In microfabrication, thermal oxidation is a way to produce a thin layer of oxide (usually silicon dioxide) on the surface of a wafer. The technique forces an oxidizing agent to diffuse into the wafer at high temperature and react with it. The rate of oxide growth is often predicted by the Deal-Grove model}}		

\newglossaryentry{deposition}{
	name=deposition,
	description={The act and technique of applying a thin film to a surface is thin-film deposition. ``Thin'' is a relative term, but most deposition techniques control layer thicknesses within a few tens of nanometers. Molecular beam epitaxy allows a single layer of atoms to be deposited at a time}}		

\newglossaryentry{cvd}{
	name=chemical vapour deposition,
	description={Chemical vapor deposition (CVD) is a chemical process used to produce high-purity, high-performance solid materials. The process is often used in the semiconductor industry to produce thin films. In a typical CVD process, the \gls{wafer} (\gls{substrate)} is exposed to one or more volatile precursors, which react and/or decompose on the \gls{substrate} surface to produce the desired deposit. Frequently, volatile by-products are also produced, which are removed by gas flow through the reaction chamber}}		

\newglossaryentry{epitaxy}{
	name=epitaxy,
	description={Epitaxy refers to the deposition of an overlayer on a crystalline \gls{substrate}, where the overlayer is in registry with the \gls{substrate}. The overlayer is called an epitaxial film or epitaxial layer. The term epitaxy comes from the Greek roots epi, meaning ``above'', and taxis, meaning ``in ordered manner''}}		

\newglossaryentry{sputtering}{
	name=sputtering,
	description={Sputtering is a process whereby atoms are ejected from a solid target material due to bombardment of the target by energetic particles. It only happens when the kinetic energy of the incoming particles is much higher than conventional thermal energies ($\gg 1\,\eV$). This process can lead, during prolonged ion or plasma bombardment of a material, to significant erosion of materials, and can thus be harmful. On the other hand, it is commonly utilised for thin-film \gls{deposition}, \gls{etching} and analytical techniques }}		

\newglossaryentry{photolithography}{
	name=photolithography,
	description={Photolithography, also termed optical lithography or \gls{UV} lithography, is a process used in microfabrication to pattern parts of a thin film or the \gls{bulk} of a \gls{substrate}. It uses light to transfer a geometric pattern from a photomask to a light-sensitive chemical ``\gls{photo resist}'', or simply ``resist'', on the \gls{substrate}. A series of chemical treatments then either engraves the exposure pattern into, or enables deposition of a new material in the desired pattern upon, the material underneath the \gls{photo resist}. For example, in complex integrated circuits, a modern \gls{CMOS} \gls{wafer} will go through the photolithographic cycle up to 50 times}}		

\newglossaryentry{photo mask}{
	name=photo mask,
	description={A photo mask is an opaque plate with holes or transparencies that allow light to shine through in a defined pattern. They are commonly used in \gls{photolithography}}}		

\newglossaryentry{photo resist}{
	name=photo resist,
	description={A photo resist is a light-sensitive material used in several industrial processes, such as \gls{photolithography} and photoengraving to form a patterned coating on a surface. A positive resist is a type of photo resist in which the portion of the photo resist that is exposed to light becomes soluble to the photo resist developer. The portion of the photo resist that is unexposed remains insoluble to the photo resist developer. A negative resist behaves in the opposite way}}		

\newglossaryentry{etching}{
	name=etching,
	description={Etching is used in microfabrication to chemically remove layers from the surface of a \gls{wafer} during manufacturing. Etching is a critically important process module, and every \gls{wafer} undergoes many etching steps before it is complete. For many etch steps, part of the \gls{wafer} is protected from the etchant by a ``masking'' material which resists etching. In some cases, the masking material is a \gls{photo resist} which has been patterned using \gls{photolithography}. Other situations require a more durable mask, such as silicon nitride}}		

\newglossaryentry{selectivity}{
	name=selectivity,
	description={If the \gls{etching} is intended to make a cavity in a material, the depth of the cavity may be controlled approximately using the etching time and the known etch rate. More often, though, \gls{etching} must entirely remove the top layer of a multilayer structure, without damaging the underlying or masking layers. The etching system's ability to do this depends on the ratio of etch rates in the two materials, and is called ``selectivity''}}		

\newglossaryentry{field oxide}{
	name=field oxide,
	description={The field oxide is the thick oxide used for masking in the \glslink{photolithography}{photolithographic} process sequence}}		

\newglossaryentry{coupling oxide}{
	name=coupling oxide,
	description={The coupling oxide (also called ``readout oxide'') is the thin oxide layer between the implanted strip and its metallisation, which isolates the readout electronics from the silicon \gls{bulk}. Its purpose is to remove the \gls{DC} component of the \gls{dark current} to avoid saturation of the input channels of the readout chip}}		

\newglossaryentry{poly head}{
	name=polysilicon head,
	description={The \gls{polysilicon} head is a special implantation at the ends of the \gls{bias resistor} to avoid a \gls{SC}}}		

\newglossaryentry{conductivity}{
	name=conductivity,
	description={Electrical conductivity or specific conductance is the reciprocal of electrical \gls{resistivity}, and measures a material's ability to conduct an electric current}}		

\newglossaryentry{p-spray}{
	name=p-spray,
	description={The p-spray blocking is a method to interrupt the \gls{accumulation} layer or \gls{inversion} layer forming on the \gls{n-side} of \glspl{sms}. The whole \gls{wafer} side is \gls{p-type} implanted with low fluence and high energy, which embeds the \gls{n-type} strips in local \gls{p-type} material. The occurring \gls{pnjunction} interrupts the \gls{electron} layer, which otherwise would short any \gls{n-type} regions present at this surface}}		

\newglossaryentry{dicing line}{
	name=dicing line,
	description={The dicing line is a structure prepared to avoid splintering during \gls{dicing}. In an \gls{IC} design, one has to make sure that there is no oxide along the path of the dicing. Some manufacturers also include a metallisation of the dicing line}}		

\newglossaryentry{fiducial circle}{
	name=fiducial circle,
	description={The fiducial circle is an imaginary circular line on a \gls{wafer}, which divides the reliable inner part from the outermost edge, where the homogeneity of the material cannot be guaranteed}}		

\newglossaryentry{sheet resistance}{
	name=sheet resistance,
	description={Sheet resistance is a measure of resistance of thin films that are nominally uniform in thickness. It is commonly used to characterise materials made by semiconductor \gls{doping}, metal \gls{deposition}, resistive paste printing, and glass coating. The total resistance $R$ of a film is calculated as $R = \frac{\rho}{t}\frac{L}{W} = R_\mathrm{s}\frac{L}{W}$, where $\rho$ is the \gls{resistivity}, $t$, $L$ and $W$ are thickness, length and width if the film, and $R_\mathrm{s}$ is the sheet resistance. A commonly used unit for the sheet resistance is $\Ohm/\square$ (read Ohms per square). For a quadratic film we can write $L=W$ and therefore $R=R_\mathrm{s}$. The resistance value $R$ of a line-shaped film can be expressed using the number of squares, regardless of the actual size of the squares}}		

\newglossaryentry{strip scan}{
	name=strip scan,
	description={A strip scan is an electrical characterisation method for \glspl{sms}. A number of measurements is carried out on every strip, yielding thorough insight into the performance and quality of the sensor. Strip scans are performed using fully automated ``probe stations''. One can either have stationary probe needles, moving the sensor to make contact to the strips, or one can have automatically moving probe needles. The latter is more versatile, but only available through commercial vendors, and expensive}}		

\newglossaryentry{common}{
	name=common,
	description={For the common \gls{p-stop} pattern, the \gls{n-type} doped strips are surrounded by a \gls{p-type} doped area covering the whole sensor. Only small regions around the strips are left unimplanted. This interrupts the \gls{accumulation} layer which hence cannot short the \gls{n-type} strips. The \gls{p-stop} implant itself is on the same (floating) potential all over the sensor and could therefore distribute any charge introduced in the implant}}		

\newglossaryentry{combined}{
	name=combined,
	description={For the combined \gls{p-stop} pattern, the \gls{n-type} doped strips are surrounded by \gls{atoll} \gls{p-stop} implants. In addition to that, both strip and \gls{atoll} implants are embedded in a \gls{p-type} doped area covering the whole sensor, like in case of the \gls{common} pattern. Only small regions around the strip and the \gls{atoll} are left unimplanted. This pattern tries to combine the benefits of the \gls{common} pattern and the \gls{atoll} pattern}}		

\newglossaryentry{pinhole}{
	name=pinhole,
	description={A pinhole is a conductive connection between the strip implant and the aluminium strip above it, which renders the \gls{cac} ineffective. A pinhole causes the \gls{DC} portion of the strip's \gls{leakage current} to flow into the input amplifier of the readout chip. The \gls{apv} chip can withstand up to seven connected pinholes, beyond that either the whole chip is dysfunctional, or the strips in question have to be disconnected}}
	
\newglossaryentry{vetronit}{
	name=Vetronit,
	description={Vetronit (also called FR4 or G10) is a glass fabric with epoxide, silicon, polyamide, melamine and phenol resin, often applied in the production of \glspl{PCB}. It is used for \gls{beam test} module frames because it is non-conductive and radiation hard}}
	
\newglossaryentry{n-sub pad}{
	name=n-sub pad,
	description={The n-sub pad allows contact to the (in our case \gls{n-type}) \gls{substrate} of the sensor. It is located in the \gls{edge ring} of the \gls{p-side}. This pad allows to bias the sensor when only the \gls{p-side} is accessible, because the n-sub pad connects to the \gls{n-side} through the \gls{bulk}}}		

\newglossaryentry{interstitial}{
	name=interstitial,
	description={An interstitial is an unbound atom in a crystal lattice, which has been knocked out of its lattice site. It is usually generated together with a \gls{vacancy}. The interstitial is mobile and can recombine with a \gls{vacancy}, disappear at the edge of the material, or be caught by other (stable) defects}}		

\newglossaryentry{vacancy}{
	name=vacancy,
	plural=vacancies,
	description={A vacancy is an empty crystal lattice site. It is usually generated together with an \gls{interstitial}. The vacancy is mobile and can recombine with an \gls{interstitial}, disappear at the edge of the material, or be caught by other (stable) defects}}		

\newglossaryentry{type inversion}{
	name=type inversion,
	description={Type inversion is an effect occurring in heavily irradiated \gls{n-type} materials. The radiation damage to the \gls{bulk} makes the \gls{n-type} material more and more \gls{p-type}-like, eventually reversing its polarity. This naturally can't happen in a \gls{p-type} material}}

\newacronym{FBK-CMM}{FBK-CMM}{FBK-CMM: Fondazione Bruno Kessler - Center for Materials and Microsystems}
\newacronym{TCT}{TCT}{TCT: Transient Current Technique}
\newacronym{VERA}{VERA}{VERA: Vienna Environmental Research Accelerator, a research apparatus of the University of Vienna, Austria}
\newacronym{IMB-CNM-CSIC}{IMB-CNM-CSIC}{IMB-CNM-CSIC: Institute of Microelectronics of Barcelona, Centre Nacional de Microelectrónica, Consejo Superior de Investigaciones Científicas}
\newacronym{MPG}{MPG}{MPG: Max-Planck-Gesellschaft zur Förderung der Wissenschaften (Germany, engl: Max Planck Society for the Advancement of Science)}
\newacronym{HLL}{HLL}{HLL: HalbLeiterLabor (Munich, Germany, engl: Semiconductor laboratory)}
\newacronym{IMSIL}{IMSIL}{IMSIL: IMplant and Sputter sImuLator (earlier, ``IMSIL'' was considered an abbreviation of IMplant simulator for SILicon technology, but IMSIL has evolved since then)}
\newacronym{PIN}{PIN}{PIN: \gls{p-type} \gls{intrinsic} \gls{n-type}}
\newacronym{UFSD}{UFSD}{UFSD: Ultra Fast Silicon Detector}
\newacronym{CCD}{CCD}{CCD: Charge-Coupled Device}
\newacronym{HEP}{HEP}{HEP: High Energy Physics}
\newacronym{SiPM}{SiPM}{SiPM: Silicon PhotoMultiplier}
\newacronym{MCP}{MCP}{MCP: MicroChannel Plate}
\newacronym{PET}{PET}{PET: \Gls{positron} Emission Tomography}
\newacronym{FET}{FET}{FET: Field Effect Transistor}
\newacronym{SDD}{SDD}{SDD: Silicon Drift Detector}
\newacronym{APD}{APD}{APD: Avalanche PhotoDiode}
\newacronym{ELAD}{ELAD}{ELAD: Enhanced LAteral Drift}
\newacronym{UV}{UV}{UV: Ultra Violet}
\newacronym{UHV}{UHV}{UHV: Ultra High Vacuum}
\newacronym{IDE}{IDE}{IDE: Integrated Development Environment}
\newacronym{GEANT4}{GEANT4}{GEANT4: GEometry ANd Tracking version 4, a platform for the simulation of the passage of particles through matter using \gls{MC} methods}
\newacronym{SRIM}{SRIM}{SRIM: Stopping and Range of Ions in Matter, a group of computer programs which calculate interaction of ions with matter}
\newacronym{PERKEO}{PERKEO}{PERKEO: to the best of our knowledge this is a proper name, and no acronym}
\newacronym{pLGAD}{\mbox{pLGAD}}{pLGAD: \gls{proton} \glslink{LGAD}{Low Gain Avalanche Detector}}
\newacronym{iLGAD}{iLGAD}{iLGAD: inverted \glslink{LGAD}{Low Gain Avalanche Detector}}
\newacronym{NEG}{NEG}{NEG: Non-Evaporable Getter}
\newacronym{TU}{TU}{TU: Technische Universität (engl.\ University of Technology)}
\newacronym{MC}{MC}{MC: \gls{Monte Carlo}}
\newacronym{UCNB}{UCNB}{UCNB: Ultra Cold \Glspl{neutron} \glssymbol{B}}
\newacronym{UCNA}{UCNA}{UCNA: Ultra Cold \Glspl{neutron} \glssymbol{A}}
\newacronym{Nab}{N$ab$}{N\glssymbol{a}\glssymbol{b}: \Gls{neutron} \glssymbol{a} \glssymbol{b}}
\newacronym{aSPECT}{$a$SPECT}{\glssymbol{a}SEPCT: \glssymbol{a} SPECTrometer}
\newacronym{aCORN}{$a$CORN}{\glssymbol{a}CORN: \glssymbol{a} CORrelation in \Gls{neutron} decay}
\newacronym{NMR}{NMR}{NMR: Nuclear Magnetic Resonance}
\newacronym{LGAD}{LGAD}{LGAD: Low Gain Avalanche Detector}
\newacronym{TR}{TR}{TR: Trans Regio}
\newacronym{SFB}{SFB}{SFB: Spezialforschungsbereich}
\newacronym{carbs}{CaRBS}{CaRBS: Calibration tools and R$\times$B Spectroscopy of \gls{neutron} decay}
\newacronym{ANR}{ANR}{ANR: Agence Nationale de la Recherche}
\newacronym{NFG}{NFG}{NFG: New Frontier Group}
\newacronym{FWF}{FWF}{FWF: Fonds zur Förderung der wissenschaftlichen Forschung (Austrian Science Fonds)}
\newacronym{OEAW}{ÖAW}{ÖAW: Österreichische Akademie der Wissenschaften (Austrian Academy of Sciences)}
\newacronym{ILL}{ILL}{ILL: Institut Laue Langevin}
\newacronym{CKM}{CKM}{CKM: Cabbibo-Kobayashi-Masukawa}
\newacronym{SMI}{SMI}{SMI: Stefan Meyer Institute for Subatomic Physics}
\newacronym{nomos}{\nomos}{\nomos: \Gls{neutron} decay prOducts Momentum Spectrometer}
\newacronym{PERC}{PERC}{PERC: \Gls{proton} \Gls{electron} Radiation Channel}
\newacronym{ANNI}{ANNI}{ANNI: A pulsed cold \gls{neutron} beam facility for particle physics at the \gls{ESS}}
\newacronym{ESS}{ESS}{ESS: European Spallation Source}
\newacronym{CaRBS}{CaRBS}{CaRBS: Calibration tools and \rxb{} spectroscopy of \gls{neutron} decay}
\newacronym{HV}{HV}{HV: High Voltage}

\newacronym{kek}{KEK}{KEK (K\textoverline{o} Enerug\textoverline{i} Kasokuki Kenky\textoverline{u} Kik\textoverline{o}) is a japanese organization whose purpose is to operate the largest particle physics laboratory in Japan, which is located in Tsukuba of Ibaraki prefecture}	

\newacronym{kek_german}{KEK}{KEK (K\textoverline{o} Enerug\textoverline{i} Kasokuki Kenky\textoverline{u} Kik\textoverline{o}) ist eine japanische Organization, die das gr\oe{}\ss{}te Teilchenforschungslabor Japans in Tsukuba (Pr\ae{}fektur Ibaraki) betreibt}

\newacronym{ECL}{ECL}{ECL: \Gls{em} \glslink{calorimeter}{CaLorimeter}}

\newacronym{SM}{SM}{SM: \gls{Standard Model} of particle physics}

\newacronym{CP}{CP}{CP: Charge-Parity, short for \gls{charge-parity conjugation}}

\newacronym{CP_german}{CP}{CP: Charge-Parity (Ladung-Parit\ae{}t), kurz f\ue{}r Ladungs-Parit\ae{}ts-Konjugation}

\newacronym{CERN}{CERN}{CERN: Organisation (Conseil) Europ\'{e}enne pour la Recherche Nucl\'{e}aire, European Organization for Nuclear Research}

\newacronym{CERN_german}{CERN}{CERN: Organisation (Conseil) Europ\'{e}enne pour la Recherche Nucl\'{e}aire, Europ\ae{}ische Organisation f\ue{}r Kernforschung}

\newacronym{ATLAS}{ATLAS}{ATLAS: A Toroidal LHC AparatuS}

\newacronym{CMS}{CMS}{CMS: Compact \glslink{muon}{Muon} \glslink{solenoid}{Solenoid}}

\newacronym{LHC}{LHC}{LHC: Large \glslink{hadron}{Hadron} Collider}

\newacronym{SUSY}{SUSY}{SUSY: SUperSYmmetry}

\newacronym{NP}{NP}{NP: New Physics}

\newacronym{KEKB}{KEKB}{KEKB: KEK \gls{B factory}}

\newacronym{KEKB_german}{KEKB}{KEKB: KEK B-Mesonen-``Fabrik''}

\newacronym{SuperKEKB}{SuperKEKB}{SuperKEKB: Super \gls{kek} \gls{B factory}, the upgrade of \gls{KEKB}}

\newacronym{SuperKEKB_german}{SuperKEKB}{SuperKEKB: Super KEK B-Mesonen-``Fabrik'', die Verbesserung von \gls{KEKB_german}}

\newacronym{PCA}{PCA}{PCA: Point of Closest Approach}

\newacronym{LDT}{LDT}{LDT: LiC Detector Toy, where LiC stands for ``Linear Collider''}

\newacronym{RAVE}{RAVE}{RAVE: Reconstruction Algorithms in Versatile Environments}

\newacronym{HER}{HER}{HER: High Energy Ring}

\newacronym{LER}{LER}{LER: Low Energy Ring}

\newacronym{IP}{IP}{IP: Interaction Point}

\newacronym{LINAC}{LINAC}{LINAC: LINear ACcelerator}

\newacronym{SVD}{SVD}{SVD: Silicon Vertex Detector}

\newacronym{CDC}{CDC}{CDC: Central Drift Chamber}

\newacronym{ACC}{ACC}{ACC: silica-Aerogel Cherenkov Counter}

\newacronym{TOF}{TOF}{TOF: Time Of Flight}

\newacronym{KLM}{KLM}{KLM: \glslink{kaon}{K$_\mathrm{L}$} and \Gls{muon} detection system}

\newacronym{DEPFET}{DEPFET}{DEPFET: DEpleted P-channel \glslink{FET}{Field Effect Transistor}}

\newacronym{PXD}{PXD}{PXD: PiXel Detector}

\newacronym{DAQ}{DAQ}{DAQ: Data AcQuisition}

\newacronym{DSSD}{DSSD}{DSSD: Double-Sided \glslink{sms}{Silicon Detector}}

\newacronym{TOP}{TOP}{TOP: Time Of Propagation}

\newacronym{RICH}{RICH}{RICH: Ring Imaging \glslink{Cherenkow counter}{Cherenkov Counter}}

\newacronym{ARICH}{ARICH}{ARICH: Aerogel Ring Imaging \glslink{Cherenkow counter}{Cherenkov Counter}}

\newacronym{IC}{IC}{IC: Integrated Circuit}

\newacronym{FEA}{FEA}{FEA: Finite Elements Analysis}

\newacronym{PCB}{PCB}{PCB: Printed Circuit Board}

\newacronym{CMOS}{CMOS}{CMOS: Complementary Metal-Oxide-Semiconductor, a technology for constructing \glslink{IC}{integrated circuits}}

\newacronym{FIFO}{FIFO}{FIFO: First In First Out, a kind of low-level memory array}

\newacronym{APSP}{APSP}{APSP: Analog Pulse Shape Processor}

\newacronym{SNR}{SNR}{SNR: \glslink{snr}{Signal-to-Noise Ratio}}

\newacronym{AC}{AC}{AC: Alternating Current}

\newacronym{DC}{DC}{DC: Direct Current}

\newacronym{FADC}{FADC}{FADC: Flash Analog-to-Digital-Converter}

\newacronym{ADC}{ADC}{ADC: Analog-to-Digital-Converter}

\newacronym{VME}{VME}{VME bus: Versa Module Eurocard-bus}

\newacronym{FPGA}{FPGA}{FPGA: Field Programmable Gate Array}

\newacronym{FTB}{FTB}{FTB: Finesse Transmitter Board}

\newacronym{COPPER}{COPPER}{COPPER: COmmon Pipeline Platform for Electronics Readout}

\newacronym{CPU}{CPU}{CPU: Central Processing Unit}

\newacronym{CO2}{CO$_2$}{\CO2{}: Carbon Dioxide}

\newacronym{MIP}{MIP}{MIP: \glslink{mip}{Minimum Ionising Particle, a relativistic particle which exhibits the minimum energy deposit per length in the Bethe-Bloch formula. This minimum occurs at an particle energy in the order of $E = 3 M c^2$, where $M$ is the particle's mass.}}

\newacronym{MOS}{MOS}{MOS: Metal-Oxide-Semiconductor}

\newacronym{sio2}{SiO$_2$}{SiO$_2$: Silicon Dioxide}

\newacronym{MPV}{MPV}{MPV: \glslink{mpv}{Most Probable Value}}

\newacronym{CCE}{CCE}{CCE: \glslink{cce}{Charge Collection Efficiency, i.e., the fraction of measurable charge carriers w.r.t.\ the number of created ones.}}

\newacronym{COG}{COG}{COG: Center-Of-Gravity} 

\newacronym{ENC}{ENC}{ENC: \glslink{enc}{Equivalent Noise Charge}}

\newacronym{FOXFET}{FOXFET}{FOXFET: Field OXide Field Effect Transistor}

\newacronym{MOSFET}{MOSFET}{MOSFET: \glslink{MOS}{Metal-Oxide-Semiconductor} Field Effect Transistor}

\newacronym{ppb}{ppb}{ppb: parts per billion}

\newacronym{EGS}{EGS}{EGS: Electronic Grade Silicon}

\newacronym{CZ}{CZ}{CZ: CZochralski}

\newacronym{FZ}{FZ}{FZ: Float Zone}

\newacronym{RF}{RF}{RF: Radio Frequency}

\newacronym{HTO}{HTO}{HTO: High Temperature Oxidation}

\newacronym{BPSG}{BPSG}{BPSG: BoroPhosphoSilicate Glass}

\newacronym{CVD}{CVD}{CVD: \glslink{cvd}{Chemical Vapour Deposition}}

\newacronym{APCVD}{APCVD}{APCVD: Atmospheric Pressure \glslink{cvd}{Chemical Vapour Deposition}}

\newacronym{LPCVD}{LPCVD}{LPCVD: Low Pressure \glslink{cvd}{Chemical Vapour Deposition}}

\newacronym{PECVD}{PECVD}{PECVD: Plasma Enhanced \glslink{cvd}{Chemical Vapour Deposition}}

\newacronym{CAD}{CAD}{CAD: Computer Aided Design}

\newacronym{RPM}{RPM}{RPM: Revolutions Per Minute}

\newacronym{AMPLE}{AMPLE}{AMPLE: Advanced Multi-Purpose LanguagE}

\newacronym{siddata}{SiDDaTA}{SiDDaTA: Silicon Detector Design and Teststructures using \gls{AMPLE}}

\newacronym{NDA}{NDA}{NDA: Non Disclosure Agreement}

\newacronym{GCD}{GCD}{GCD: Gate Controlled Diode}

\newacronym{SRP}{SRP}{SRP: Spreading Resistance Profiling}

\newacronym{QTC}{QTC}{QTC: Quality Test Center}

\newacronym{HEPHY}{HEPHY}{HEPHY: Institute of High Energy PHYsics}

\newacronym{SMU}{SMU}{SMU: Source Measurement Unit}

\newacronym{LCR}{LCR}{LCR: Inductance (L), capacitance (C), resistance (R)}

\newacronym{GPIB}{GPIB}{GPIB: General Purpose Interface Bus}

\newacronym{SPS}{SPS}{SPS: Super \Gls{proton} Synchrotron}

\newacronym{PDF}{PDF}{PDF: Probability Density Function}

\newacronym{RMS}{RMS}{RMS: Root Mean Square}

\newacronym{PVC}{PVC}{PVC: PolyVinylChloride}

\newacronym{DUT}{DUT}{DUT: Device Under Test}

\newacronym{sckcen}{SCK$\bullet$CEN}{SCK$\bullet$CEN: StudieCentrum voor Kernenergie $\bullet$ Centre d'Etude de l'\'energie Nucl\'eaire}

\newacronym{RITA}{RITA}{RITA: Radio Isotope Test Arrangement}

\newacronym{BRIGITTE}{BRIGITTE}{BRIGITTE: Big Radius Installation under Gamma Irradiation for Tailoring and Testing Experiments}

\newacronym{EUDET}{EUDET}{EUDET: Detector R\&D towards the \glslink{ILC}{International Linear Collider}, a project supported by the European Union in the 6$^{th}$ Framework Programme (FP6) structuring the European Research Area}

\newacronym{ILC}{ILC}{ILC: International Linear Collider}

\newacronym{CLW}{CLW}{CLW: CLuster Width}

\newacronym{MAMBO}{MAMBO}{MAMBO: Mother (MAMa) BOard}

\newacronym{REBO}{REBO}{REBO: REpeater BOard}

\newacronym{DATCON}{DATCON}{DATCON: DATa CONcentrator}

\newacronym{NIM}{NIM}{NIM: Nuclear Instrumentation Module}

\newacronym{TLU}{TLU}{TLU: Trigger Logic Unit}

\newacronym{TRHX}{TRHX}{TRHX: Temperature and Relative Humidity eXtended~\cite{trhx}}

\newacronym{RnD}{R\&D}{R\&D: Research and Development}

\newacronym{IISS}{IISS}{IISS: Institute for Integrated Sensor Systems}

\newacronym{MAD}{MAD}{MAD: Median Absolute Deviation}

\newacronym{LED}{LED}{LED: Light Emitting Diode}

\usepackage[capitalize,nameinlink,noabbrev]{cleveref}

\usepackage{placeins}
\usepackage{amssymb}
\usepackage{subcaption}
\usepackage{upgreek}
\usepackage[figuresright]{rotating}

\usepackage{enumitem}
\setitemize{noitemsep,topsep=0pt,parsep=0pt,partopsep=0pt}

\journal{Nucl.\ Instrum.\ Methods Phys.\ Res.\ Section A}

\begin{document}

\begin{frontmatter}

\title{First Results for the pLGAD Sensor for Low-Penetrating Particles}


\cortext[mycorrespondingauthor]{Corresponding author}
\author[smiaddress]{Waleed Khalid\corref{mycorrespondingauthor}}
\ead{waleed.khalid@oeaw.ac.at}

\author[smiaddress]{Manfred Valentan}
\author[cnmaddress]{Albert Doblas}
\author[cnmaddress]{David Flores}
\author[cnmaddress]{Salvador Hidalgo}
\author[smiaddress,tuaddress]{Gertrud Konrad}
\author[smiaddress]{Johann Marton}
\author[cnmaddress]{Neil Moffat}
\author[smiaddress]{Daniel Moser}
\author[smiaddress]{Sebastian Onder}
\author[cnmaddress]{Giulio Pellegrini\corref{mycorrespondingauthor}}
\ead{giulio.pellegrini@csic.es}
\author[cnmaddress]{Jairo Villegas}



\address[smiaddress]{Stefan Meyer Institute for Subatomic Physics, Austrian Academy of Sciences, Kegelgasse 27, 1030 Vienna, Austria}
\address[cnmaddress]{Centro Nacional de Microelectrónica, IMB-CNM-CSIC, 08193 Cerdanyola del Vallès, Barcelona, Spain}
\address[tuaddress]{Atominstitut, TU Wien, Stadionallee 2, 1020 Vienna, Austria}

\begin{abstract}
	
	Silicon sensors are the go-to technology for high-precision sensors in particle physics.
	But only recently low-noise silicon sensors with internal amplification became available. 
	The so-called Low Gain Avalanche Detector (LGAD) sensors have been developed for applications in High Energy Physics, but lack two characteristics needed for the measurement of low-energy protons ($<\SI{60}{keV}$):
	a thin entrance window (in the order of tens of \si{\nm}) and the efficient amplification of signals created near the sensor’s surface (in a depth below \SI{1}{\micro\metre}).
	
	In this paper we present the so-called proton Low Gain Avalanche Detector (pLGAD) sensor concept and some results from characterization of the first prototypes of the sensor. The pLGAD is specifically designed to detect low-energy protons, and other low-penetrating particles.
	It will have a higher detection efficiency than non-silicon technologies, and promises to be a lot cheaper and easier to operate than competing silicon technologies.
	
\end{abstract}

\begin{keyword}
precision physics \sep low-energy \sep high efficiency \sep internal gain \sep LGAD  \sep pLGAD  
\end{keyword}

\end{frontmatter}


\section{Introduction}
	\label{sec:introduausschlagenction}
	Low-energy precision physics experiments investigate particles that have a low penetration depth ($<\SI{1}{\micro\meter}$) within a detection material due to their low momenta. This fact demands specialized detection systems as the Signal to Noise Ratio (SNR) of a conventional High Energy Physics (HEP) detector is quite high to offer a good separation of the signal from the noise, e.g. for the case of $aSPECT$ \cite{martin_postacc} or N$ab$ \cite{Broussard2019} experiments. While there are detection techniques available that can be utilized for the detection of low penetrating particle such as, DEpleted Field Effect Transistors (DEPFET), Silicon Drift Detectors (SDDs) etc., they often are expensive, complicated to operate, or require cooling. As a consequence, their application is often difficult if not unfeasible for small scale experiments. 

A very promising solution to the aforementioned problem is the usage of the LGAD (Low Gain Avalanche Detector) or iLGAD (inverted Low Gain Avalanche Detector) technology. However, as the these detectors are designed for HEP applications, specific changes have to be made before such a detector can be used for the detection of low penetrating particles.

\subsection{The challenge of the entrance window}
		\label{sec:entrance_window}
		When attempting to detect the signal of a low-penetrating particle, a sensor faces three distinct requirements: thin non-active layers, relatively high Charge Collection Efficiency (CCE) at the surface (as close to 1 as possible), and a readout with high SNR. 
A low-energy particle loses a portion of its energy within the non-active layers of the sensor and another near the surface of the silicon substrate that has a higher defect density than the bulk, and is usually heavily doped to form a field stop. Therefore, for any technology to function well for the detection of such particles, a thin entrance window (combination of the non-active layers and the heavily doped layer of silicon with reduced CCE) is essential.

For the NoMoS (Neutron decay prOducts MOmentum Spectrometer) measurement concept~\cite{Mos19}, the benchmark is a proton with \SI{15}{\kilo\eV}. In this case, the maximum depth in which energy is deposited is roughly \SI{300}{\nano\metre}, assuming a sensor with a thin metal passivation layer of \SI{4}{\nano\meter} aluminium oxide and an incidence angle of the impinging proton of \SI{0}{\degree} (simulated 500,000 protons using IMSIL \cite{hobler_monte_1995,hobler_random_2006}). IMSIL, which is binary collision Monte Carlo simulation software, was used to study the track of protons within such a detector geometry without inclusion of the electric field of the detector. Using these simulations, estimations of total number of primary \glspl{ehpair} generated as a function of depth within the detector was obtained.

The CCE, which describes the probability that an \gls{ehpair} survives and can be measured, is usually significantly smaller than 1 near the surface of the sensor. Therefore the combination of the signal with the CCE of the sensor, leads to a total signal that is significantly lower than that of a MIP (Minimum Ionizing Particle) traversing a standard \SI{300}{\micro\meter} silicon sensor. For the NoMoS benchmark proton, approximately 4000 \glspl{ehpair} (dividing the mean electronic energy deposited by the protons by \SI{3.6}{\electronvolt}) are produced within the sensor, compared to the 24,000 \glspl{ehpair} from a MIP.

\section{A new silicon sensor concept}
		\label{sec:concept}

\subsection{The pLGAD Concept}
\label{sec:plgad}

\begin{figure}[t]

	\centering
	\includegraphics[width=0.7\columnwidth]{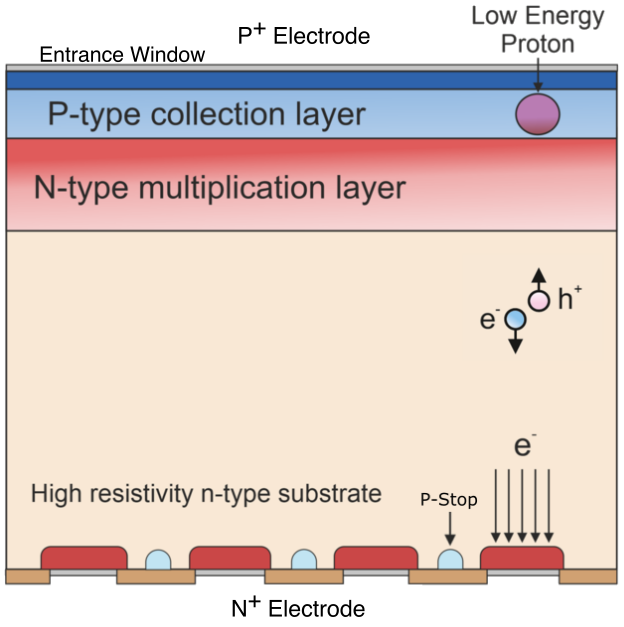}

	\caption{In contrast to the iLGAD concept \cite{Pellegrini2016}, the pLGAD concept uses inverted implantation polarity.
		This way, only signal electrons created in the collection region are amplified, directly next to the entrance window.
		The main contribution to the signal stems from the secondary electrons. Darker colors represent higher doping concentration.}
	\label{fig:ilgad_plgad}
\end{figure}

Based on the iLGAD concept \cite{Pellegrini2016}, the proposed proton Low Gain Avalanche Diode (pLGAD) sensor concept \cite{plgad_patent} is designed with inverted doping compared to a traditional iLGAD, and takes matters two steps further: the pLGAD is equipped with a very thin, unstructured passivation layer of \SI{4}{\nm} of aluminium oxide (or wishfully, a thin \SI{15}{\nm} aluminium conduction layer) and a thin field stop in the order of $\SI{50}{\nano\meter}$ level. The sensor concept furthermore introduces a collection region by having the \gls{pnjunction} and the multiplication layer deeper in the bulk, away from the entrance window. Lastly, the polarity of the signal-collecting electrode is chosen to be an N-type, so that the signal electrons drift to the readout sensor side and cross the multiplication layer, as illustrated in Fig. \ref{fig:ilgad_plgad}. This also has the added benefit that thermal \glspl{ehpair} produced within the bulk of the sensor are not multiplied as holes do not cause impact ionization when crossing the N-type multiplication layer.

In principle, the collection region defines the uniformity of the gain seen by any particle. Its depth can be adjusted by placing the gain region deeper into the sensor (via epitaxial growth) or near the unstructured entrance window (via ion implantation) during the production stage. This gives pLGAD an additional advantage of being flexible for low-energy, high precision experiments. 

In the pLGAD design, only signals created close to the entrance window are amplified. This makes the sensor uninteresting for HEP applications, but perfectly suited for low-energy physics experiments.

In summary, the pLGAD concept makes clever use of two systematic asymmetries to detect low-penetrating particles: the signal created only close to the surface, and the fact that only electrons are multiplied.
\subsection{Simulation of doping concentrations and breakdown voltage}
In order to obtain the appropriate breakdown voltage, $V_{bd}$, and the expected gain in the pLGAD, numerical simulations using TCAD Sentaurus \cite{tcad} were performed.

\begin{figure}[hbt!]
	
		\centering
		\includegraphics[width=0.83\columnwidth]{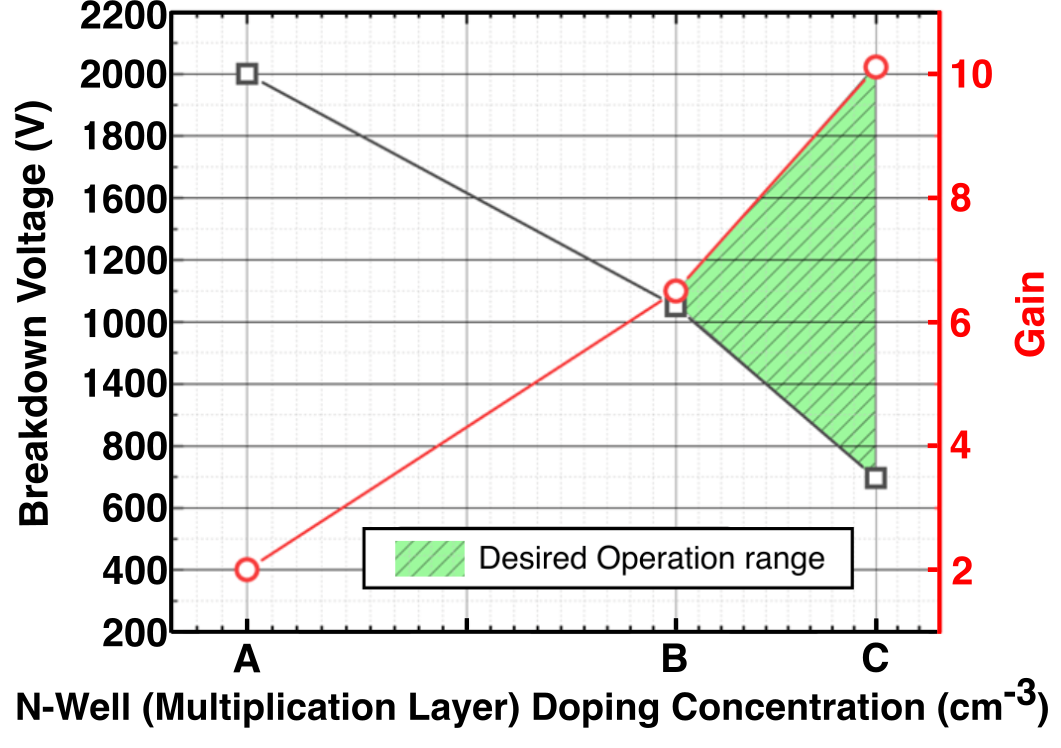}
		\caption{TCAD simulations of the expected gain and the breakdown voltage of a pLGAD sensor as a function of three different doping concentrations (represented by A, B, and C, of the multiplication N layer). The operating range refers to the desired breakdown voltage and gain.}
		\label{dopingvbd}
		

\end{figure}

The key quantity for fine-tuning the operation of the sensor is the doping concentration of the multiplication layer (N-Well) as shown in Fig. \ref{dopingvbd}. The green area in the plot represents the operating range of interest for the sensor. This area was chosen to have the lowest $V_{bd}$ while obtaining an acceptable gain for the detection of the low-energy protons.
  From the preliminary TCAD simulations (Fig. \ref{dopingvbd}), it could be concluded that a pLGAD with a gain of 10 and a $V_{bd}$ of $700\,\si{\volt}$ can be fabricated by using an N-well doping concentration, that is represented by C in Fig. \ref{dopingvbd}.

\section{Characterization of first prototypes}
\label{sec:prototypes}
\begin{figure}[hbt!]
			\begin{subfigure}{\columnwidth}	

		\centering
		\includegraphics[width=0.8\columnwidth]{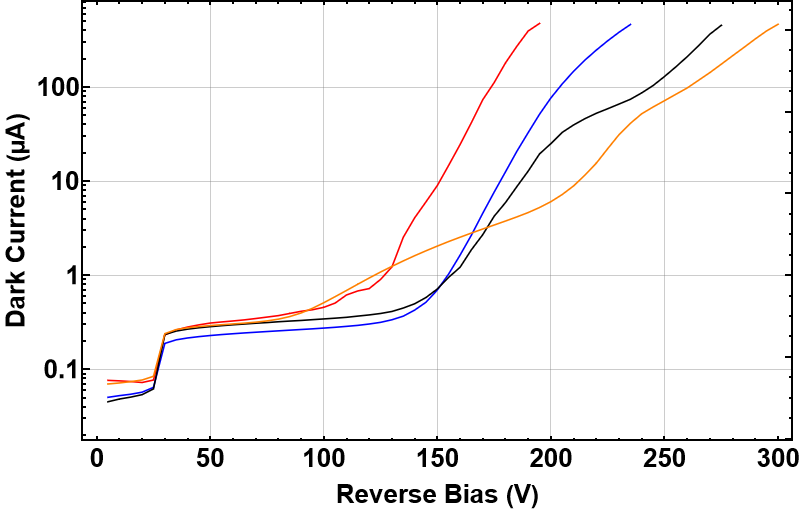}
		\caption{The I-V curves show a uniform operating range of the diodes for $\SI{30}{\volt}<V_{bias}<\SI{100}{\volt}$.}
		\label{fig:plgad_iv}
				\end{subfigure}
			\begin{subfigure}{\columnwidth}
			\centering
			\includegraphics[width=0.8\columnwidth]{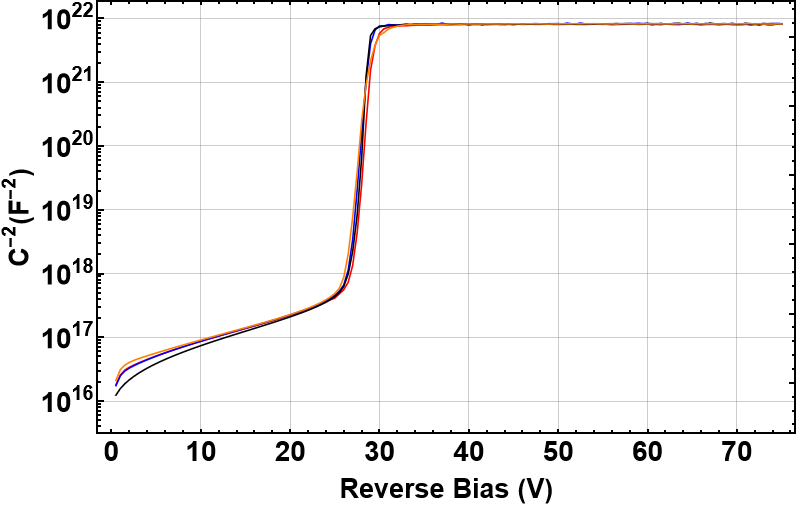}
			\caption{The C-V curves show that full depletion of the diodes is reached at $V_{bias}>\SI{30}{\volt}$.}
			\label{fig:plgad_cv}
		\end{subfigure}
	\caption{Measured dark current and capacitance-voltage values of the four prototype diodes.}
	\label{fig:plgad_cv_iv}
	
\end{figure}

In order to verify that the proposed sensor concept is feasible, a first production run of pLGAD sensors without a P-type collection region, a standard passivation layer of silicon nitrate albeit with a thin backplane, and shallow multiplication implant was conducted at IMB-CNM-CSIC. Four $\SI{5.3x5.3}{}\,$\SI{}{\milli\meter^2} diodes from this run were chosen for initial characterization, and their current-voltage (I-V) and capacitance-voltage (C-V) curves measured, as shown in Figures \ref{fig:plgad_iv} and \ref{fig:plgad_cv}. From these measurements, it was concluded that the diodes reach a full depletion at $V_{bias}>\SI{30}{\volt}$. 

\begin{figure}[hbt!]

		\centering
		\includegraphics[width=0.8\columnwidth]{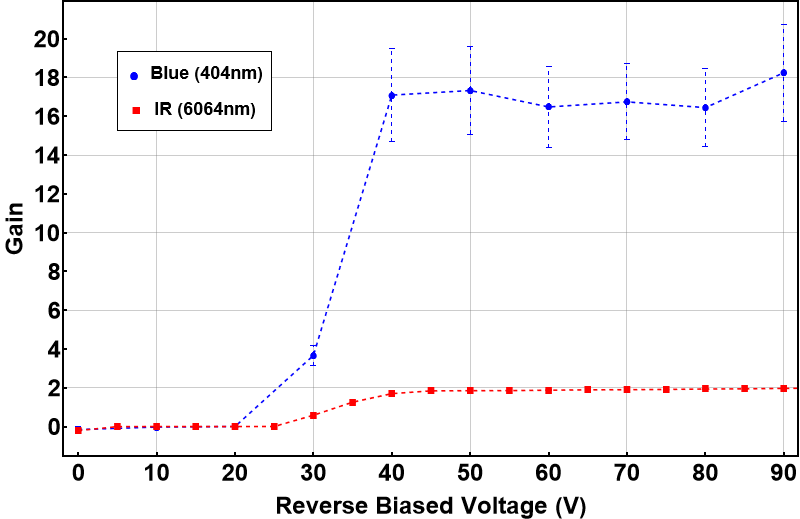}
		\caption{Measured gain of one prototype diode obtained by using a blue and an infrared laser. For details see text.}
		\label{fig:gainplot}
	
\end{figure}

In order to determine the gain of the prototypes, voltage scan for one of the diodes was performed in a TCT (Transient Current Technique) setup with a blue laser of $\SI{404}{\nano\meter}$, and an infrared laser of $\SI{1064}{\nano\meter}$. Figure \ref{fig:gainplot} shows that the diode, as opposed to a PIN diode under the same conditions, exhibits a higher gain factor for the blue laser as compared to infrared one. As the blue light (absorption length of $<\SI{1}{\micro\meter}$ \cite{silicon_wavelength_absorption}) is totally absorbed within the gain layer, it sees a higher gain compared to the infrared light (absorption length of $\approx\SI{0.1}{\centi\meter}$ \cite{silicon_wavelength_absorption}). This is a direct consequence of the polarity chosen for the pLGAD concept, as the majority of the \glspl{ehpair} from the infrared laser are deposited within the bulk of the sensor where the majority charge carriers (electrons) drift away from the multiplication layer. The large errors bars for the blue light in Fig. \ref{fig:gainplot} are due to the low signal to noise ratio for the PIN diode when measured using the TCT setup. The error bars can be reduced by bump bonding the PIN to a newer PCB and retaking measurements but are sufficient for first impressions. The complete characterization results will be published soon.

%
%
%
%
%
%
%

\section{Possible applications}
\label{sec:applications}

The pLGAD sensor concept allows low-energy precision physics experiments to benefit from all advantages of silicon sensors without having to sacrifice detection efficiency or spatial resolution. Its thin \gls{entrance window} and internal amplification are desired in low-energy precision physics experiments. It can easily be adapted to the needs of different experiments, e.g. for the case of low-energy protons in NoMoS \cite{Mos19}, $a$SPECT \cite{martin_postacc}, or N$ab$ \cite{Broussard2019}.

A pLGAD sensor can be used especially but not exclusively for low-energy spectroscopy experiments with high energy resolution. The sensor works not only for \glspl{proton}, but also for alpha particles, low-energy ions, and with a suitable conversion layer coated on the \gls{entrance window} for soft X-rays and \glspl{neutron}. Furthermore, the technology can also be used for the monitoring of low-energy beam lines, for Time of Flight (TOF) experiments with a low light yield such as TOF-PET and, also for space-borne experiments.

\section{Summary and Outlook}
		\label{sec:conclusions}

We presented a novel silicon sensor concept and selected results from its first proof-of-principle production run. The new technology is especially geared towards low-penetrating particles, with a range of below \SI{1}{\micro\metre} in silicon (see \cref{sec:entrance_window}).

The pLGAD sensor concept is a polarity-inverted iLGAD, in which only \glspl{ehpair} created close to the \gls{entrance window} are multiplied (see \cref{sec:concept}). Due to this the \gls{leakage current} and its corresponding noise remains unamplified and the pLGAD behaves like a simple, \gls{planar} sensor for high energy particles, while it offers full amplification for low-penetrating particles.
%

The preliminary results from the characterization of the first production run look fairly promising. A gain of 17 is obtained from a blue laser of wavelength $\SI{404}{\nano\meter}$ compared to a gain of 2 for an infrared laser of $\SI{1064}{\nano\meter}$. Since the multiplication layer can be placed at any thickness within the sensor, future production runs will include a collection region of thickness $>0$ and a thinner passivation or a metallic conduction layer.
These steps will be taken as our driving application for the sensor was the NoMoS measurement concept~\cite{Mos19}, a new method of momentum-spectroscopy for the charged decay products  from \gls{neutron} \gls{beta decay}, which requires position-resolved \gls{proton} detection on a large-area sensor, at a reasonable cost.
Any kind of low-penetrating particles can be detected using a pLGAD sensor, like soft X-rays, charged ions, alpha particles and with a thin conversion layer, also neutrons.

\subsection{Acknowledgments}

Many thanks go to Gerhard Hobler (TU Wien, Vienna, Austria) for discussions in setting up the IMSIL simulations.

The IMSIL simulations were performed using the CLIP Cluster (Vienna, Austria) \cite{cluster}. This work is partially supported by the Austrian Academy of Sciences within the New Frontiers Groups Programme NFP 2013/09, the Austrian Science Fund under contract No. W1252 (DK-PI), TU Wien (Vienna, Austria) and the SMI (Vienna, Austria). This work is also partially supported by the Spanish Ministry of Science and Innovation through the Particle Physics National Program (FPA2017-85155-C4-2-R and RTI2018-094906-B-C22) and by the European Union FEDER funding program.


\bibliography{references}


\end{document}